\documentclass[final,a4paper]{article}

\usepackage[hyphens]{url}
\usepackage[top=2cm,bottom=3cm,left=2.4cm,right=2.4cm]{geometry}
\usepackage{mathtools,amssymb}
\usepackage[usenames,dvipsnames,table]{xcolor}
\usepackage[titletoc]{appendix}
\usepackage{mathtools,amssymb}
\usepackage[makeroom]{cancel}
\usepackage{graphicx}
\usepackage{epstopdf}
\usepackage{mathrsfs}
\usepackage{enumitem}
\usepackage{stackrel}
\usepackage{float}
\usepackage{booktabs}
\usepackage[normalem]{ulem}  
\usepackage{cancel} 
\usepackage{hyperref}
\usepackage[color]{showkeys} 
\usepackage{longtable}
\usepackage{physics}
\usepackage{empheq}
\usepackage{cite} 

\DeclareMathAlphabet{\mathpzc}{OT1}{pzc}{m}{it}

\allowdisplaybreaks
\allowbreak

\definecolor{refkey}{gray}{0.75}
\definecolor{labelkey}{RGB}{155,48,48}
\renewcommand*\showkeyslabelformat[1]{%
  \fbox{\parbox[t]{0.8\marginparwidth}{\raggedright\normalfont\scriptsize\url{#1}}}}

\usepackage{etoolbox}
\makeatletter
\patchcmd{\hyper@makecurrent}{%
    \ifx\Hy@param\Hy@chapterstring
    \let\Hy@param\Hy@chapapp
    \fi
}{%
    \iftoggle{inappendix}{
	\@checkappendixparam{chapter}%
	\@checkappendixparam{section}%
	\@checkappendixparam{subsection}%
	\@checkappendixparam{subsubsection}%
	\@checkappendixparam{paragraph}%
	\@checkappendixparam{subparagraph}%
    }{}%
}{}{ \errmessage{failed to patch}}

\newcommand*{\@checkappendixparam}[1]{%
	\def\@checkappendixparamtmp{#1}%
	\ifx\Hy@param\@checkappendixparamtmp
	\let\Hy@param\Hy@appendixstring
	\fi
}
\makeatletter

\newtoggle{inappendix}
\togglefalse{inappendix}

\apptocmd{\appendix}{\toggletrue{inappendix}}{}{\errmessage{failed to patch}}
\apptocmd{\subappendices}{\toggletrue{inappendix}}{}{\errmessage{failed to patch}}

\newcommand{\lsim}{\mathrel{\hbox{\rlap{\lower .55ex
\hbox{$\sim$}} \kern-.3em \raise.4ex \hbox{$<$}}}}
\newcommand{\gsim}{\mathrel{\hbox{\rlap{\lower.55ex
\hbox{$\sim$}} \kern-.3em \raise.4ex \hbox{$<$}}}}

\begin{document}


\newcommand{\partiald}[2]{\dfrac{\partial #1}{\partial #2}}
\newcommand{\be}{\begin{equation}}
\newcommand{\ee}{\end{equation}}
\newcommand{\f}{\frac}
\newcommand{\s}{\sqrt}
\newcommand{\lm}{\mathcal{L}}
\newcommand{\wm}{\mathcal{W}}
\newcommand{\om}{\mathcal{O}_{n}}
\newcommand{\ep}{\epsilon}

\def\gap#1{\vspace{#1 ex}}
\def\del{\partial}
\def\eq#1{(\ref{#1})}
\def\fig#1{Fig \ref{#1}} 
\def\re#1{{\bf #1}}
\def\bull{$\bullet$}
\def\nn{\nonumber}
\def\ub{\underbar}
\def\nl{\hfill\break}
\def\ni{\noindent}
\def\bibi{\bibitem}
\def\vev#1{\langle #1 \rangle} 
\def\mattwo#1#2#3#4{\left(\begin{array}{cc}#1&#2\\#3&#4\end{array}\right)} 
\def\tgen#1{T^{#1}}
\def\half{\frac12}
\def\floor#1{{\lfloor #1 \rfloor}}
\def\ceil#1{{\lceil #1 \rceil}}

\def\Tr{{\rm Tr}}

\def\mysec#1{\gap1\ni{\bf #1}\gap1}
\def\mycap#1{\begin{quote}{\footnotesize #1}\end{quote}}

\def\Red#1{{\color{red}#1}}

\def\Om{\Omega}
\def\a{\alpha}
\def\b{\beta}
\def\l{\lambda}
\def\g{\gamma}
\def\e{\epsilon}
\def\Si{\Sigma}
\def\p{\phi}
\def\z{\zeta}

\def\lan{\langle}
\def\ran{\rangle}

\def\bit{\begin{item}}
\def\eit{\end{item}}
\def\benu{ \begin{enumerate} }
\def\eenu{ \end{enumerate} }

\def\tr{{\rm tr}}
\def\intk#1{{\int\kern-#1pt}}


\parindent=0pt
\parskip = 10pt

\def\al{\alpha}
\def\ga{\gamma}
\def\Ga{\Gamma}
\def\G{\Gamma}
\def\be{\beta}
\def\de{\delta}
\def\De{\Delta}
\def\ep{\epsilon}
\def\ro{\rho}
\def\la{\lambda}
\def\La{\Lambda}
\def\ka{\kappa}
\def\om{\omega}
\def\si{\sigma}
\def\th{\theta}
\def\ze{\zeta}
\def\ne{\eta}
\def\del{\partial}
\def\cdev{\nabla}

\def\gh{\hat{g}}
\def\Rh{\hat{R}}
\def\Boxh{\hat{\Box}}
\def\Kb{\mathcal{K}}
\def\phit{\tilde{\phi}}
\def\gt{\tilde{g}}
\newcommand{\h}{\hat}
\newcommand{\ti}{\tilde}
\newcommand{\sD}{{\mathcal{D}}}
\newcommand{\colored}[1]{ {\color{turquoise} #1 } }
\newcommand{\propBbd}{\mathcal{G}}
\newcommand{\propBB}{\mathbb{G}}
\newcommand{\christof}[3]{ {\Ga^{#1}}_{#2 #3}}
\def\ads{AdS$_{\text{2}}$~}
\def\GN{G$_{\text{N}}$~}
\def\zb{{\bar{z}}}
\def\fb{\bar{f}}
\def\delb{\bar{\del}}
\def\wb{\bar{w}}
\def\gb{\bar{g}}
\def\gp{g_+}
\def\gm{g_-}
\def\phit{\tilde{\phi}}
\def\mut{\tilde{\mu}}
\def\xb{\bar{x}}
\def\yb{\bar{y}}
\def\xp{x_+}
\def\xm{x_-}
\def\finv{\mathfrak{f}_i}
\def\fbinv{\bar{\mathfrak{f}}_i}
\def\gc{\mathfrak{g}}
\def\gcb{\bar{ \mathfrak{g}}}
\def\disc{\mathcal{D}}
\def\rhp{\mathbb{H}}
\def\picklemma{Schwarz-Pick lemma}
\def\mobius{M\"{o}bius~}
\def\ft{\tilde{f}}
\def\zet{\tilde{\ze}}
\def\taut{\tilde{\tau}}
\def\thet{\tilde{\theta}}
\def\slr{\ensuremath{\mathbb{SL}(2,\mathbb{R})}}
\def\slc{\ensuremath{\mathbb{SL}(2,\mathbb{C})}}
\def\nh{\hat{n}}
\def\cD{\mathcal{D}}
\def\Lfg{\mathfrak{f}}

\renewcommand{\real}{\mathbb{R}}

\newcommand*{\Cdot}[1][1.25]{%
  \mathpalette{\CdotAux{#1}}\cdot%
}
\newdimen\CdotAxis
\newcommand*{\CdotAux}[3]{%
    {%
	\settoheight\CdotAxis{$#2\vcenter{}$}%
	\sbox0{%
	    \raisebox\CdotAxis{%
		\scalebox{#1}{%
		    \raisebox{-\CdotAxis}{%
			$\mathsurround=0pt #2#3$%
		    }%
		}%
	    }%
	}%
	\dp0=0pt %
	\sbox2{$#2\bullet$}%
	\ifdim\ht2<\ht0 %
	\ht0=\ht2 %
	\fi
	\sbox2{$\mathsurround=0pt #2#3$}%
	\hbox to \wd2{\hss\usebox{0}\hss}%
    }%
}

\newcommand\hcancel[2][black]{\setbox0=\hbox{$#2$}%
\rlap{\raisebox{.45\ht0}{\textcolor{#1}{\rule{\wd0}{1pt}}}}#2}

\renewcommand{\arraystretch}{2.5}%
\renewcommand{\floatpagefraction}{.8}%

\def\newthing{\marginpar{{\color{red}****}}}
\reversemarginpar
\def\bz{{\bar z}}
\def\by{{\bar y}}
\def\bw{{\bar w}}
\def\delb{{\bar \del}}
\def\bep{{\bar \epsilon}}
\def\bfa{{\bf a}}
\def\bfb{{\bf b}}
\def\bfc{{\bf c}}
\def\bfd{{\bf d}}

\hypersetup{pageanchor=false}
\begin{titlepage}
    \begin{flushright}
	 ICTS/2017/01, TIFR/TH/16-28
    \end{flushright}

    \vspace{.4cm}
    \begin{center}
	\noindent{\Large \bf{Coadjoint orbit action of Virasoro group
            and two-dimensional quantum gravity dual to SYK/tensor
            models}}\\
	\vspace{1cm}
	Gautam Mandal$^a$\footnote{mandal@theory.tifr.res.in},
	Pranjal Nayak$^a$\footnote{pranjal@theory.tifr.res.in}
	and Spenta R. Wadia$^b$\footnote{spenta.wadia@icts.res.in} 

	\vspace{.5cm}
	\begin{center}
	    {\it a. Department of Theoretical Physics}\\
	    {\it Tata Institute of Fundamental Research, Mumbai 400005, 
	    India.}\\
	    \vspace{.5cm}
	    {\it b. International Centre for Theoretical Sciences}\\
	    {\it Tata Institute of Fundamental Research, Shivakote,
	    Bengaluru 560089, India.}
	\end{center}

	\gap2


    \end{center}

    \begin{abstract}

The Nambu-Goldstone (NG) bosons of the SYK model are described by a
coset space \emph{Diff}/SL(2,R), where \emph{Diff}, or {\it Virasoro
  group}, is the group of diffeomorphisms of the time coordinate
valued on the real line or a circle. It is known that the coadjoint
orbit action of \emph{Diff} naturally turns out to be the two-dimensional
quantum gravity action of Polyakov without cosmological
constant, in a certain gauge, in an asymptotically flat
spacetime. Motivated by this observation, we explore Polyakov action
with cosmological constant and boundary terms, and 
study the possibility of such a two-dimensional quantum gravity model being the AdS dual to the low
energy (NG) sector of the SYK model. We find strong evidences for this duality: (a) the bulk action admits an exact family
of asymptotically AdS$_2$ spacetimes, parameterized by
\emph{Diff}/SL(2,R), in addition to a fixed conformal factor of a simple
functional form; (b) the bulk path integral reduces to a path integral
over \emph{Diff}/SL(2,R) with a Schwarzian action; (c) the low temperature
free energy qualitatively agrees with that of the SYK model. We show,
up to quadratic order, how to couple an infinite series of bulk
scalars to the Polyakov model and show that it reproduces the coupling
of the higher modes of the SYK model with the NG bosons.

\end{abstract}

\begin{figure}[H]
\begin{minipage}{0.45\linewidth}
\centerline{\includegraphics[height=2.5cm]{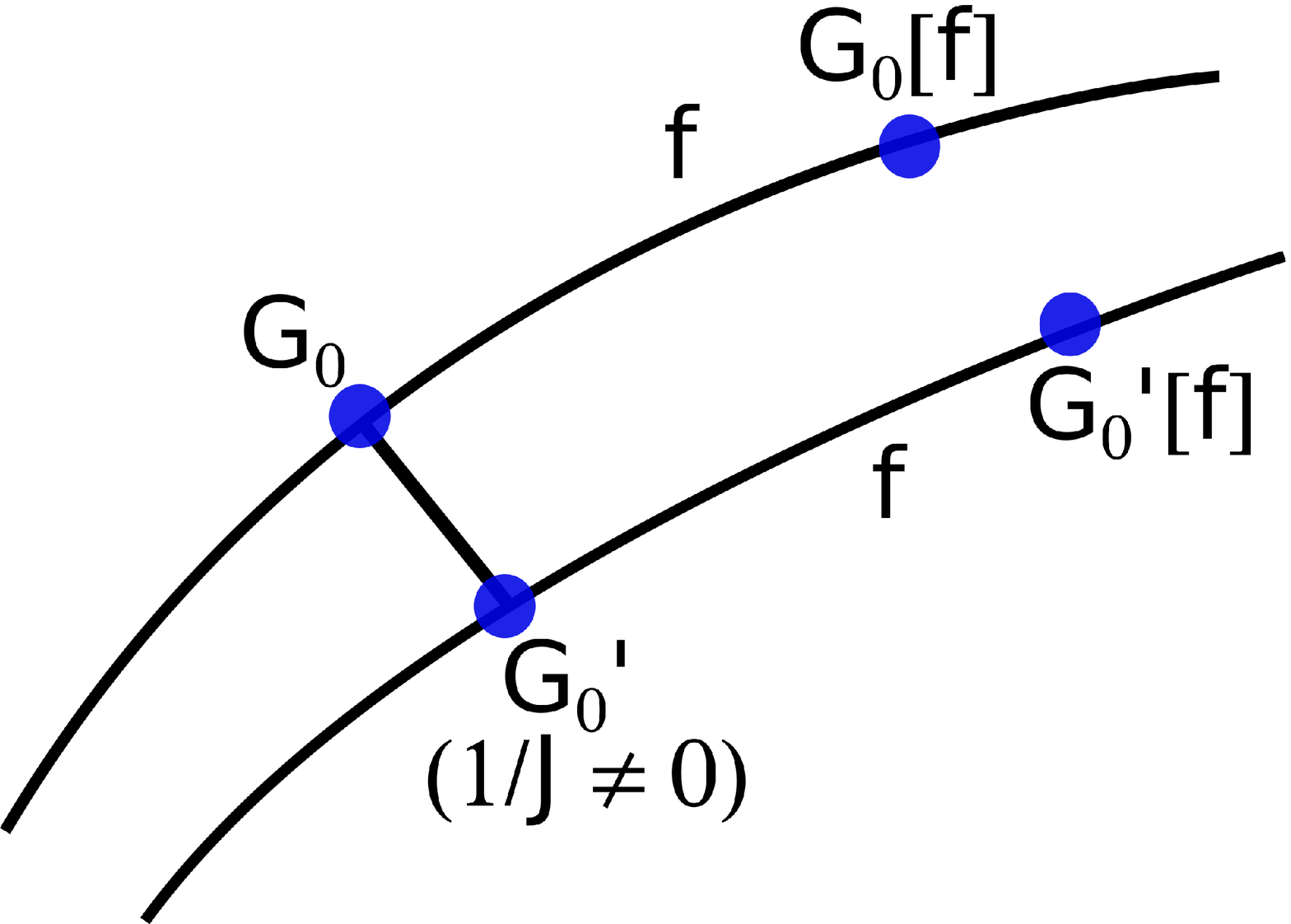}}
\end{minipage}
\boldmath$\longleftrightarrow$
\begin{minipage}{0.45\linewidth}
\centerline{\includegraphics[height=2.5cm]{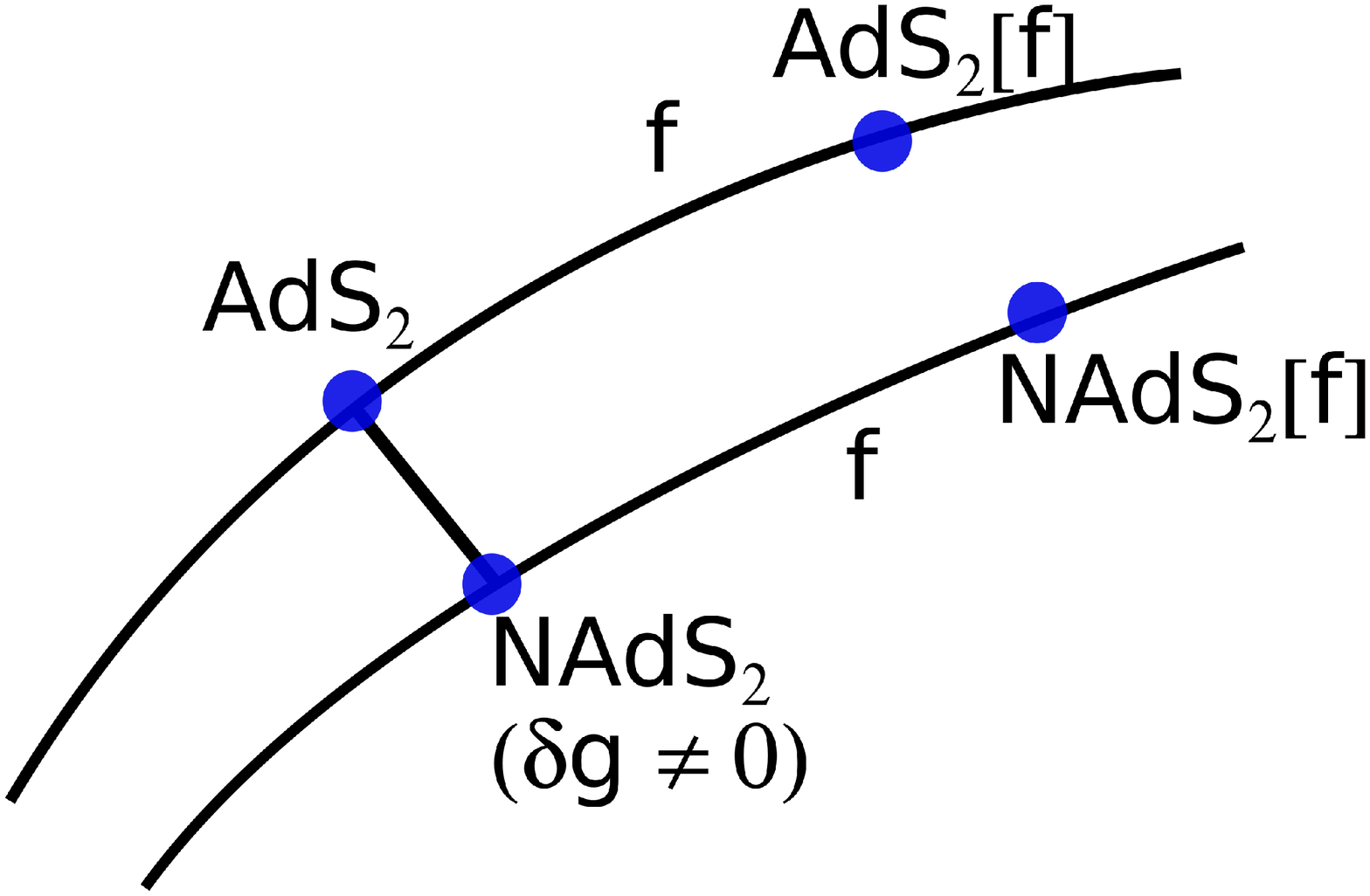}}
\end{minipage}
\end{figure}

\end{titlepage}

\pagenumbering{roman}
\tableofcontents
\pagenumbering{arabic}
\setcounter{page}{1}

\section{Introduction and Summary}\label{sec:Intro}
The Sachdev-Ye-Kitaev (SYK) model and other tensor models that have
universal IR properties \cite{Kitaev-talks:2015, Sachdev:1992fk,
  Sachdev:2010um, PhysRevX.5.041025, Maldacena:2016hyu, Gurau:2016lzk,
  Witten:2016iux, Klebanov:2016xxf}, are quantum mechanical models of
large $N$ fermionic particles, described by a Hamiltonian which, for
Euclidean time $\tau=i t$, can be viewed alternatively as a
one-dimensional statistical model of fermions. The SYK model has
random couplings $J_{i_1 i_2...i_q}$, representing disorder, and does
not correspond to a unitary quantum mechanics. A different version
without the random disorder, but with the same leading large $N$
behaviour, has been proposed by Gurau \cite{Gurau:2009tw,
  Gurau:2016lzk}, Witten \cite{Witten:2016iux}, and Klebanov and
Tarnopolsky \cite{Klebanov:2016xxf}. Here we are interested only in the
large $N$ behaviour and will call the set of models SYK-type
models. More recently, higher dimensional generalizations of such
models have also been a subject of study with the expectation that
various interesting properties that make such models a good playground
to study black hole physics can be carried over to the higher
dimensions, \cite{Gu:2016oyy,Gross:2016kjj,Berkooz:2016cvq}.

SYK-type models have drawn a lot of attention in the literature
recently (see, \cite{Kitaev-talks:2015, Maldacena:2016hyu,
  Gurau:2016lzk, Witten:2016iux,
  Klebanov:2016xxf,Berkooz:2016cvq,Gu:2016oyy,Engelsoy:2016xyb,
  Turiaci:2016cvo, Jevicki:2016ito, Jevicki:2016bwu,
  Gross:2016kjj,Fu:2016vas,Cotler:2016fpe,Garcia-Garcia:2016mno,Davison:2016ngz,Krishnan:2016bvg,Li:2017hdt,Cvetic:2016eiv}
for a partial list of related developments), primarily because of the
following features in a large $N$ limit:

(1) There is an infrared fixed point with an emergent time
reparametrisation symmetry, denoted henceforth as \emph{Diff}.\footnote{We use \emph{Diff} to denote either Diff($R$) or Diff($S^1$),
  depending on whether we are at zero temperature or finite
  temperature. This group is alternatively called the {\it Virasoro}
  group.\label{ftnt:diff1}} The symmetry is spontaneously broken, at
  the IR fixed point, to \slr~ by the large $N$ classical solution,
  leading to Nambu-Goldstone (NG) bosons characterized by the coset
  \emph{Diff}/\slr.\footnote{As explained later in more detail, unlike
    in higher dimensions where Nambu-Goldstone modes are zero modes of
    the action promoted to spacetime fields, here they remain zero
    modes (do not acquire kinetic terms) since they cannot be made
    dependent on any other dimension.} At the IR fixed point all these
  are precise zero modes of the action as one might expect from a
  one-dimensional CFT. Slightly away from the IR fixed point, the
  \emph{Diff} symmetry is explicitly broken, the `Nambu-Goldstone'
modes cease
  to be zero modes and their dynamics is described by a Schwarzian
  term (which is the equivalent of a `pion mass' term). It has been
  conjectured that (see, e.g. \cite{Maldacena:2016upp}) that this
  situation is similar to a bulk model in which the AdS$_2$ symmetry
  is slightly broken (this is called a {\it near} AdS$_2$ geometry, in
   the sense of an s-wave reduction from higher dimensions, as in
  \cite{Almheiri:2014cka}).

(2) The possibility of a gravity dual is further reinforced by the
fact that the Lyapunov exponent in the SYK model saturates the chaos bound, which is
characteristic of a theory of gravity that has black hole solutions \cite{Maldacena:2015waa,Jensen:2016pah,Cvetic:2016eiv}.

(3) The full model has an approximately linearly rising (`Regge-type')
spectrum of conformal weights near the IR fixed point, with $O(1)$
anomalous dimension even for operators with spin higher than two. This
behaviour is unexpected both from string theory in the limit $\al' \to
0$, or from Vasiliev theory (see, for example,
\cite{Maldacena:2016hyu}). Thus while the dynamics of the soft modes
appears to have a simple dual gravity description, it is not clear if
it can naturally incorporate the rest of the Regge-type spectrum
description. In this paper we primarily concern ourselves with a bulk
gravity dual which describes the soft modes. We leave the larger issue
for later work.

The strategy we pursue for the proposed bulk dual is as follows. 

\begin{figure}[H]
\thispagestyle{empty}
\begin{minipage}{0.45\linewidth}
\centerline{\includegraphics[height=2.5cm]{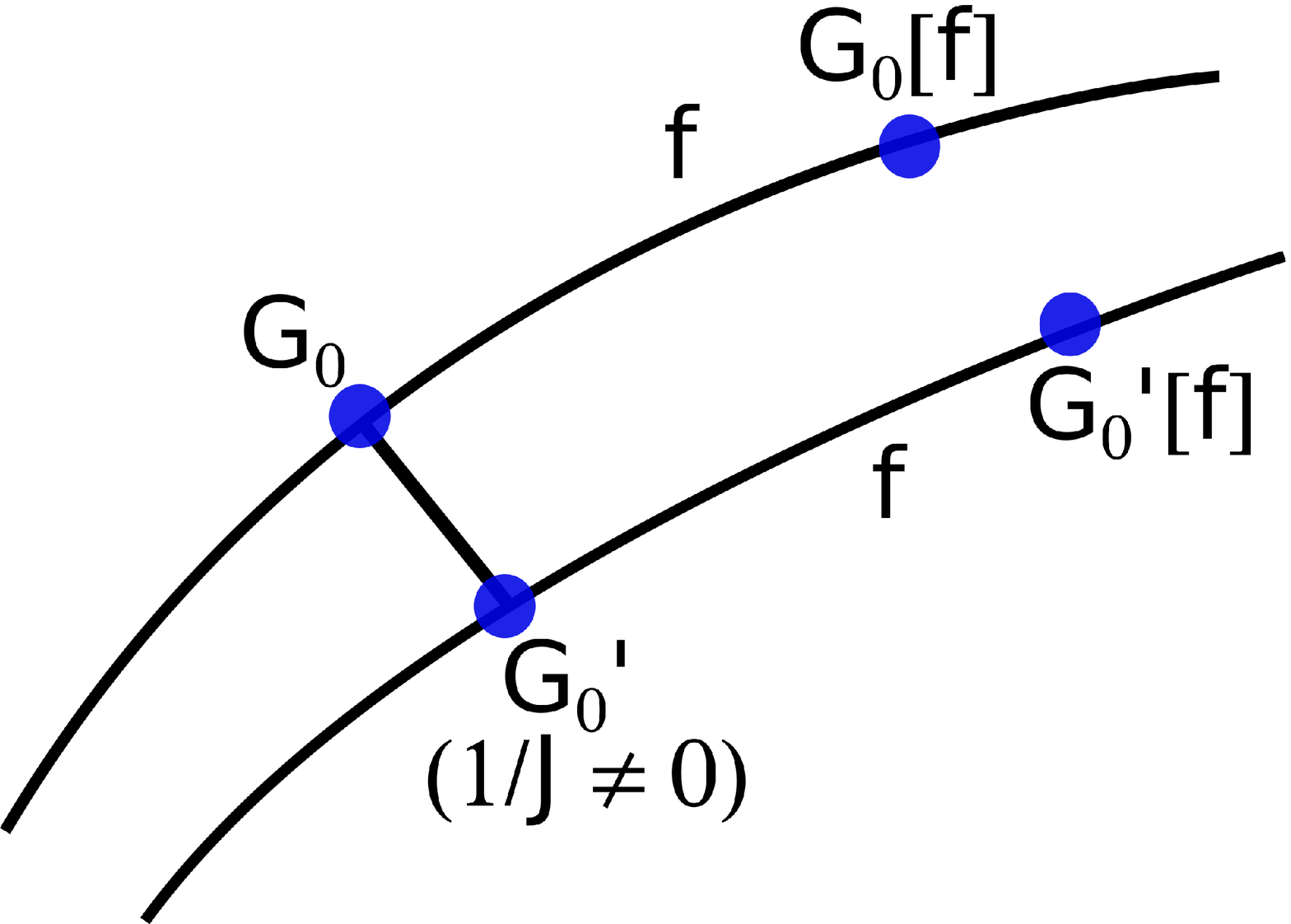}}
\end{minipage}
{\boldmath$\longleftrightarrow$}
\begin{minipage}{0.45\linewidth}
\centerline{\includegraphics[height=2.5cm]{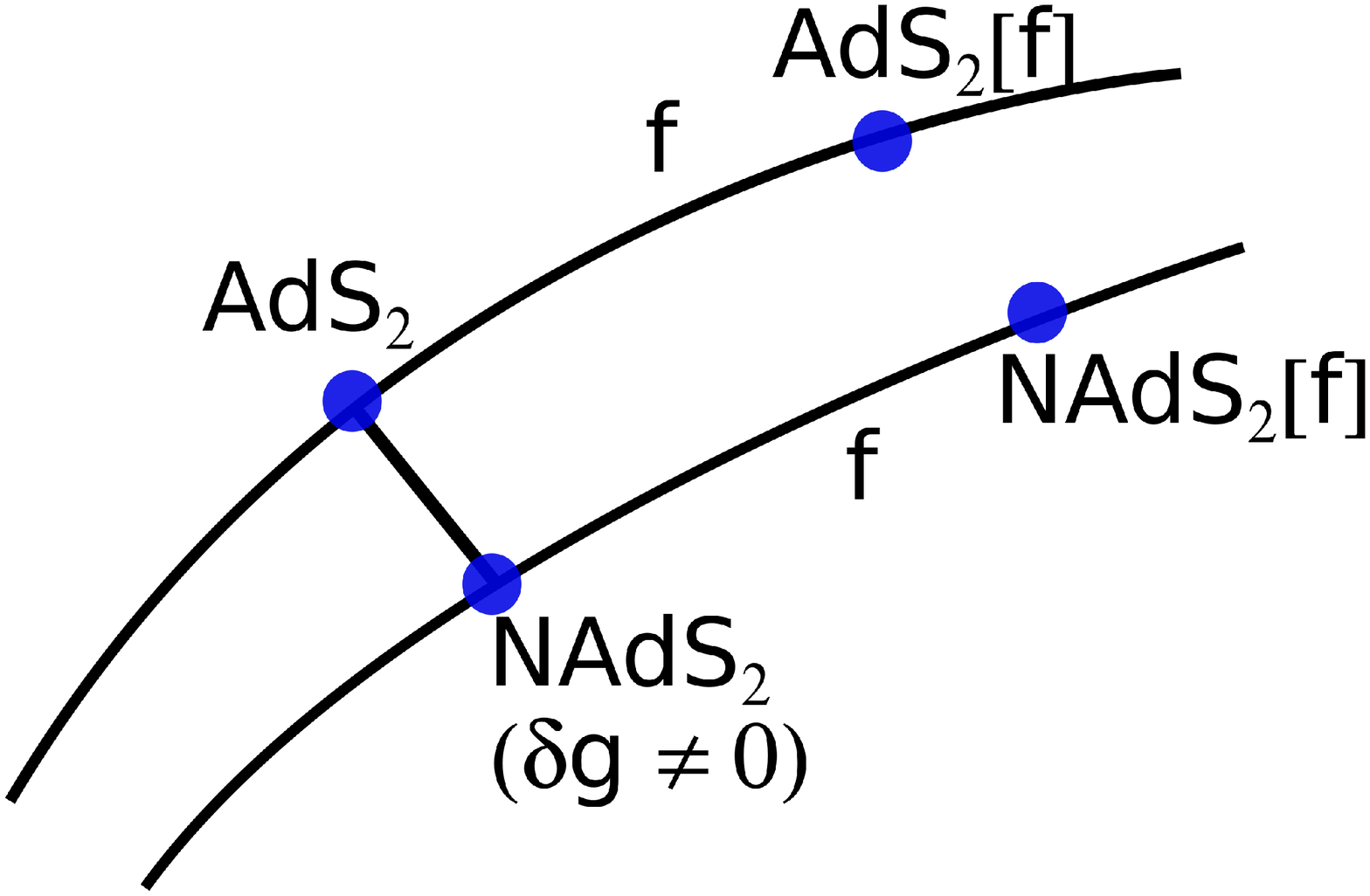}}
\end{minipage}
\caption{\footnotesize In the left panel, the top curve represents the
  Diff($R$)-orbit (or a Diff($S^1$)-orbit at finite temperature), {\it
    at the IR fixed point} $J= \infty$, of the classical large $N$
  solution for the fermion bilocal $G_0(\tau_1, \tau_2) \sim (\tau_1-
  \tau_2)^{-2 \Delta}$; this represents the Nambu-Goldstones of
  Diff($R$)/\slr~. The lower curve represents the orbit of a
  deformed solution $G_0'$ slightly away from the IR fixed point, with
  a small positive $1/J$. In the right panel, the top curve represents
  the orbit of the AdS$_2$ spacetime (these are asymptotically AdS$_2$
  spacetimes, the two-dimensional equivalent of Brown-Henneaux
  geometries, which we will describe explicitly in Section
  \ref{sec:asymp-ads}).  The bottom curve represents the orbit of a
  slightly deformed AdS$_2$ spacetime NAdS$_2$, with a controlled
  non-normalizable deformation (see \autoref{sec:asymp-ads}).}
\label{fig-coadjoint}
\end{figure}

As explained in \cite{Kitaev-talks:2015, Maldacena:2016hyu}, the NG
modes of the SYK-type model can be characterized by \emph{Diff} orbits
of the classical solution $G_0$ (at the IR fixed point $J=\infty$) or
\emph{Diff} orbits of $G_0'$ which is the deformed value of $G_0$ after
turning on a small value of $1/J$ (see figure
\ref{fig-coadjoint}). Any given point on the \emph{Diff} orbit can be
obtained from the reference point, $G_0$ or $G_0'$, by the action of  
an appropriate one-dimensional diffeomorphism.

It is shown in \cite{Alekseev:1988ce, Rai:1989js} that the space of
coadjoint orbits of \emph{Diff} can be quantized using a natural
symplectic form {\it a la} Kirillov \cite{Kirillov:1972}, leading to
Polyakov's two-dimensional quantum gravity action
\cite{Polyakov:1987zb}.  This observation is reminiscent of the
emergent two-dimensional bulk description from the $c=1$ model, which
is a matrix quantum mechanics. It was found in \cite{Dhar:1992hr,
  Dhar:1992rs} that the semiclassical (large $N$) singlet
configurations of the matrix quantum mechanics, described by fermion
droplets on a two-dimensional phase plane, could be understood as
coadjoint orbits of $W_\infty$ algebra generated by bi-local boson
operators made out of fermions. A representation of this algebra in
$c=1$ was found in \cite{Das:1991qb}. The coadjoint orbit action {\it
  a la} Kirillov \cite{Kirillov:1972} in the space of these
configurations gave rise to a two-dimensional action whose low energy
sector reproduced the (massless) tachyons of two-dimensional string
theory.\footnote{The precise correspondence required some additional
  structure (`leg-poles'); see \cite{Balthazar:2017mxh} for some
  recent insight.} A similar approach was taken in
\cite{Mandal:2005wv} to arrive at a moduli space action of LLM
geometries \cite{Lin:2004nb} describing half-BPS giant gravitons.

Following the above examples, one might wonder whether such a
two-dimensional quantum gravity action, obtained by the coadjoint
orbit method, naturally describes a bulk dual to the SYK model. It
turns out that {\it a priori} it is not possible since the gravity
action does not have a cosmological constant and it describes
asymptotically flat spaces. This prompts us to consider a
generalization of the Polyakov action, which includes a cosmological
constant and boundary terms (the boundary terms are found by requiring
the existence of a well-defined variational principle; these are also
the terms required by consistency with the Weyl anomaly in a manifold
with a boundary, see \autoref{app:Wess-Zumino} for details).
The new action, described in Section
\ref{sec:polyakov}, has asymptotically AdS$_2$ geometries as solutions
(see Section \ref{sec:asymp-ads} and \ref{sec:Indu-grav-to-Liou}),
which are all generated from AdS$_2$ by the action of \emph{Diff}. The
schematics of these solutions is described in the right panel of
Figure \ref{fig-coadjoint}.

The main point of the paper is that the {\it two-dimensional quantum
  gravity theory}, arrived at in this fashion, {\it provides a bulk
  dual to the Nambu-Goldstone sector of the SYK models}. We find a
number of strong evidences for this duality:
\begin{enumerate}
\item[(a)] the space on which path integral of the bulk theory is 
performed reduces to \emph{Diff}/\slr~, which is the same
as that of the Nambu-Goldstone bosons in the SYK model. In the bulk
theory these degrees of freedom emerge as the space of large
diffeomorphisms (analogous to Brown-Henneaux diffeomorphisms in
AdS$_3$). In addition to these, the bulk metric admits a fixed, non-dynamical conformal factor of a simple functional form. In the SYK theory this parameterizes the departure from strong coupling.

\item[(b)] The bulk path integral reduces to a path integral over \emph{Diff}/\slr~
with a Schwarzian action \autoref{sec:Hydrodynamics}, characterized by
a non-zero overall coefficient coming from the conformal factor.

\item[(c)] the low temperature free energy qualitatively agrees with that SYK model, \autoref{sec:Hydrodynamics}. In the Discussion section,
we show how to go beyond the low energy sector, and describe the higher mass modes of the SYK model, by introducing bulk matter fields. We show, up to quadratic order, how to couple an infinite series of bulk scalars to the Polyakov model and show that it reproduces the coupling of the higher modes of the SYK model with the NG bosons.
\end{enumerate}

Our paper is organized as follows. In Section \ref{sec:polyakov} we
present a motivation for our proposed bulk action \eq{eq:Liou-cov-bdy}
from the viewpoint of coadjoint orbits of \emph{Diff}/\slr. In the
two subsequent sections, we analyze the theory in the conformal gauge
$ds^2 = e^{2\phi} \widehat{ds^2}$.  In Section
\ref{sec:Indu-grav-to-Liou}, we describe solutions of the equation of
motion where $\widehat{ds^2}$ represents pure \ads\ geometry; it turns
out that the `Liouville mode' $\phi$ gets completely fixed by the
equations of motion (in fact, by just the Virasoro constraints, as
shown in Appendix \ref{app:off-shell}), up to three real parameters
which define boundary conditions for the metric. In Section
\ref{sec:asymp-ads} we find a larger class of solutions, which represent
large diffeomorphisms of \ads\ (similar to Brown-Henneaux geometries
in asymptotically AdS$_3$ spacetimes). These are normalizable modes of
the metric (`boundary gravitons') and represent dynamical variables of
the path integral, which is described in Section
\ref{subsec:proper-path}. In Section \ref{sec:Hydrodynamics} the
effective action of these boundary gravitons is obtained by an
on-shell evaluation of the path-integral; it is found to be given by a
Schwarzian \eq{schwarzian-summary}. Thus, the boundary gravitons are
found to represent the pseudo-Nambu-Goldstone modes of the SYK model.
In Section \ref{sec:Thermodynamics}, we focus on a large
diffeomorphism which leads to a Euclidean black hole geometry (this
turns the boundary direction into a circle). On-shell action for this
geometry reproduces the qualitative features of the free energy of the
SYK models. Detailed comparison with the SYK model is carried out in
Section \ref{sec:matching}. Finally, in Section \ref{sec:discussion},
we discuss how to describe the `hard' modes of the SYK model in terms
of external probe scalars coupled to the metric. The Appendices
contain detailed derivations of some formulae and supplementary
arguments.

\section{2D quantum gravity action \label{sec:polyakov}}

In this section, we briefly review some of the material on coadjoint
orbits of \emph{Diff} in \cite{Witten:1987ty, Alekseev:1988ce,
  Rai:1989js}, focussing on the emergence of 2D quantum gravity
represented by the Polyakov action \cite{Polyakov:1987zb}.

As explained in \cite{Kitaev-talks:2015, Maldacena:2016hyu}, and
briefly mentioned in the Introduction, the zero modes of the SYK model
at the IR fixed point (we suggestively call these the Nambu-Goldstone
(NG) modes, although they differ somewhat from their higher
dimensional counterpart, as explained below) are given by \emph{Diff}
transforms of the large $N$ condensate of the bilocal `meson' variable
$G(\tau_1, \tau_2) = \psi_I(\tau_1)\psi_I(\tau_2),$\footnote{We are
  using a generalized notation here, in which `$I$' denotes the
  appropriate indices of a given SYK/tensor model. For example, in SYK
  model it denotes the `flavour indices' of fermions $\psi_i$, while
  in Witten-Gurau model it denotes the tri-fundamental index the
  fermions carry.}
\begin{align}
& G_0(\tau_1, \tau_2) \sim \f1{(\tau_1 - \tau_2)^{2\Delta}}
\xrightarrow{f \in {\rm \emph{Diff}}(R^1)} G_0[f](\tau_1, \tau_2)
\nonumber\\
& G_0[f](f(\tau_1), f(\tau_2)) \equiv  G_0(\tau_1, \tau_2) 
\left(\f{\del f(\tau_1)}{\del \tau_1}
\f{\del f(\tau_2)}{\del \tau_2}\right)^{-\Delta}
\label{g0-f}
\end{align}
Here $f: \tau \to f(\tau)$ represents an element of Diff($R^1$).
This orbit is represented pictorially by the top curve in the left
panel of Fig \ref{fig-coadjoint}.  In case of finite temperature, the
time direction is considered Euclidean and compactified into a circle
of size $\b = 1/T$: in that case the appropriate group of
transformations is Diff($S^1$).

The second line of the above equation essentially says that $G$
transforms as a bilocal tensor of weight $2\Delta$ under the
diffeomorphism $f$. For later reference, we give the infinitesimal
version of this transformation as represented in the space of bilocal
variables. (for $f(\tau)= \tau + \ep(\tau)$)
\begin{align}
\delta_\ep G(\tau_1, \tau_2) 
=\left[ \Delta\left(\del_{\tau_1}\ep(\tau_1)+ \del_{\tau_2}\ep(\tau_2)
\right) +  \ep(\tau_1)\del_{\tau_1} + \ep(\tau_2)\del_{\tau_2}
\right] G(\tau_1, \tau_2) 
\label{ep-g}
\end{align}
Note that $G_0$, as defined in the first line, is
invariant under $\slr$, i.e. under \emph{Diff} elements of the form
$h(\tau)= (a\tau + b)/(c\tau + d)$, with $ad - bc=1$. This implies
that the orbit described above parameterizes a coset \emph{Diff}/$\slr$,
namely the set of \emph{Diff} elements quotiented by the identification
$f(\tau) \sim f(h(\tau))$. 

An important issue in the context of the SYK model is the quantum
mechanical realization of the \emph{Diff} algebra; in particular, it
is an important question what the central charge of the corresponding
Virasoro algebra is. We will find below, in terms of the bulk dual
described by \eq{eq:Liou-cov-bdy}, that the central charge of the
two-dimensional realization is proportional to $N$.\footnote{More
  precisely, the \emph{Diff} group is realized here as a subgroup of a
  two-dimensional conformal algebra which is unbroken by the presence
  of the boundary.}			

In higher dimensions, such as in pion physics, the elements of the
coset represent Nambu-Goldstone bosons, with kinetic terms given by a
nonlinear sigma model (see, e.g. the discussion of pions in
\cite{Weinberg:1996kr}, Chapter 19).  The Nambu-Goldstone bosons are zero-modes promoted to spacetime-dependent fields. In the SYK model, the zero-modes are described by
$f(\tau)$, or in the infinitesimal form $\ep(\tau)$, \eq{ep-g}. Their definition already uses up the only dimension available in the model, and hence they cannot be made dependent on any other coordinate and remain zero modes (do not pick any kinetic terms). As explained
in \cite{Kitaev-talks:2015, Maldacena:2016hyu}, when we move away from
the strict IR limit (i.e. when we turn on a small value of $1/J$),
these cease to be zero modes and pick up a non-zero action, given in
terms of the Schwarzian derivative
\begin{align}
S_{\rm eff} \sim \f{N}{J} \int d\tau \{f, \tau\},\; {\rm where}~~
\{f, \tau\} \equiv \f{f'''(\tau)}{f'(\tau)} - \f32 
\left(\f{f''(\tau)}{f'(\tau)}\right)^2  
\label{SYK-schwarz}
\end{align}
In spite of the appearance of the derivatives, the above is a
`potential' term for the zero modes, similar to a pion mass
term.\footnote{One way to appreciate this is to regard the Euclidean
  time as a discrete lattice and think of the `time' derivatives in
  terms of discrete differences $f'(\tau) \sim f_{i+1} - f_i$ where
  $f(\tau)$ is regarded as a collection of constant zero modes $f_i$.}

\subsection{Coadjoint orbits}\label{sec:coadjoint}
The above discussion shows that the degrees of freedom of the low
energy (NG) sector of the SYK theory are characterized by elements of
$M$= \emph{Diff}/\slr. In particular, the free energy is given by a
path integral over $M$ with the above Schwarzian action.

In this subsection, we address the question of possible quantization
of this configuration space. This question has a natural
interpretation in terms of AdS/CFT correspondence, since the bulk path
integral can, in a sense, be regarded as a radial quantum evolution of
boundary data \cite{Susskind:1998dq, Faulkner:2010jy,
  Heemskerk:2010hk}.  \cite{Susskind:1998dq, Faulkner:2010jy,
  Heemskerk:2010hk}\footnote{See \cite{Mandal:2016rmt} for a detailed
  treatment of the boundary wavefunction which represents the CFT data
  accurately.}

The quantum theory envisaged above has a configuration space given by
the group of paths in $M$ (the group of closed paths in $M$ is called
loop($M$)). An action functional on this space was formulated in
\cite{Alekseev:1988ce, Rai:1989js}, using the formalism of coadjoint
orbits and the resulting symplectic form in $M$ \cite{Kirillov:1972,
  Witten:1987ty}.  Let us consider a path ${\cal P}(\sigma)$ in the
space of \emph{Diff} elements, with ${\cal P}(0) = P_0$, ${\cal P}(1)
= P_1$. Since each point of the path is represented by a
diffeomorphism, we can label the path as $f(\tau, \sigma)$ where the
initial point $P_0$ corresponds to some diffeomorphism $f_0(\tau)$ and
the final point $P_1$ to another diffeomorphism $f_1(\tau)$. The above
mentioned action functional for such a path, also called the coadjoint
orbit action or the Kirillov action, is given by
\cite{Alekseev:1988ce, Rai:1989js} (where the symplectic form is
$\Omega = d\Theta$)
\begin{align} 
S_{\rm Kirillov} &=  
\int d\sigma  \Theta(\sigma, \{f(\tau, \sigma)\}) 
\nonumber\\
=&  \int d\sigma d\tau \left[-b_0\!\left(\{f(\tau)\}
\right) f'\dot f
+ \f{c}{48 \pi} \f{f'}{\dot f} \left(\f{\dddot f}{\dot f} -
2 \f{{\ddot f}^2}{{\dot f}^2}\right) \right]
\label{diff-kirillov}
\end{align}
where $\dot f= \del_\tau f, f'= \del_\sigma f$ etc. Here $c$
represents a possible central term in the coadjoint representation of
\emph{Diff} \cite{Witten:1987ty, Alekseev:1988ce, Rai:1989js}; $b_0$ is an
arbitrary functional, representing the choice of a reference point on
the orbit (different inequivalent orbits correspond to different
inequivalent choices of $b_0$.) 

It was observed in \cite{Alekseev:1988ce, Rai:1989js} that, with the
choice $b_0=0$ (we discuss this more later), the Kirillov action
becomes the same as the two-dimensional quantum gravity action of
Polyakov \cite{Polyakov:1987zb}
\begin{equation}\label{eq:polyakov}
    S[g] = \f{c}{24 \pi} \int_{\Ga} \sqrt{g} R \; \f1{\Box} R
\end{equation}
where the metric is \cite{Alekseev:1988ce}\footnote{The function $f(\tau, \sigma)$ here  should be compared with $F(x,t)$ of \cite{Alekseev:1988ce}}
\begin{align}
ds^2 = \del_\sigma\! f\  d\tau d\sigma
\label{alekseev-gauge}
\end{align}
Here, $R$ is the Ricci scalar of the geometry, $\frac{1}{\Box}$ is a 
notation used for the inverse of the scalar Laplacian in the geometry.

\subsection{Two-dimensional quantum gravity action}

It is rather remarkable that the two-dimensional quantum gravity
action of Polyakov emerges from the quantization of the \emph{Diff}
configuration space.\footnote{In the foregoing discussion, the fact
  that the \emph{Diff} symmetry is slightly broken does not appear to be
  taken into account. Shortly we discuss how the broken \emph{Diff} symmetry gets incorporated from the 2D gravity perspective.} Identifying such a quantization with the
holographic path integral, as mentioned in the previous subsection,
one would tend to identify \eq{eq:polyakov} with a possible bulk dual
for the Nambu-Goldstone sector of the SYK model.  This does not work,
however, since the action \eq{eq:polyakov} does not have a
cosmological constant and therefore pertains to asymptotically flat
spaces without a boundary. To qualify as the bulk dual, the classical
action must admit asymptotically AdS$_2$ spaces as solutions. Is there
a natural generalization of the Polyakov action \eq{eq:polyakov}
which admits such solutions?

It turns out that there is such an action, given by\footnote{One might wonder whether other non-local terms like $\pqty{\frac1\Box R}^n$, $n\in\mathbb{Z}^+$ are allowed in the action. It can be shown that including such higher order terms in general leads to equations of motion that do not admit an asymptotically \ads spacetime.}
 \begin{equation}\label{eq:Liou-cov-bdy}
\boxed{S_{cov}[g] = \f1{16\pi b^2} \int_{\Ga} \sqrt{g} \bqty{R \; 
\f1{\Box} R - 16 \pi \mu } + \f1{4\pi b^2} \int_{\del\Ga} \sqrt{\ga} \Kb \; 
\f1{\Box} R + \f1{4\pi b^2} \int_{\del\Ga} \sqrt{\ga} \Kb \, \f1\Box\Kb}
\end{equation}
Here $\Kb$ is the extrinsic curvature of the boundary. The constant
$b^2= \f{3}{2c}$ is the dimensionless Newton's constant in two
dimensions; we are interested in the classical limit $b\to 0$. A
bulk cosmological constant, $(-\mu)<0$\footnote{We have already incorporated a negative sign while writing the action, thus leaving $\mu>0$}, is also included (to accommodate
asymptotically AdS$_2$ spaces). The boundary terms are dictated by the
requirement of a well-defined variational principle (see
\autoref{sec:var-Poly} for derivation); these terms can also be
independently derived from the considerations of Weyl anomaly on
manifolds with a boundary, see \autoref{app:Wess-Zumino}. We have presented a discussion of the
quantum corrections contributing to the action in
\autoref{sec:quantum-corr}.

We propose that the modified quantum gravity action
\eq{eq:Liou-cov-bdy} describes a bulk dual of the low energy sector of
the SYK model. In the rest of the paper, we present strong evidence
in favour of this duality.

In the next section, we will discuss more details of the above action.
We will discuss in the subsequent section the \emph{Diff} orbit of AdS$_2$
(asymptotically AdS$_2$ metrics) in detail, and show that they are
solutions of the equations of motion. We should note that the specific
realization of this \emph{Diff} orbit will differ somewhat from that of the
above discussion. The most important difference is that in the above
discussion (which assumes spacetime without a boundary) various points
of the \emph{Diff} orbit are actually diffeomorphic in 2D; in our
construction below, the \emph{Diff} orbits involve {\it large diffeomorphisms
  in 2D} which are nontrivial near the boundary, and hence constitute
physically distinguished configurations.

It is important to emphasize the following points:

 \begin{enumerate} 

\item
  The action \eq{eq:polyakov} involves the dynamical variables
  $f(\tau,\sigma)$ representing the loop space $L(\emph{Diff})$ (more
  precisely, $L(M)$, $M=$ \emph{Diff}/$\slr$). It describes a
  quantization of $M$, which is different from simply integrating over
  $M$. The latter emerges in the description of the pseudo-Nambu-Goldstone modes
  of the SYK model. It is possible to identify
  the quantization of $M$ as the two-dimensional boundary dual to
  gravity on AdS$_3$ (see, e.g., \cite{Maloney:2007ud}).\footnote{We
    thank D.~Stanford and E.~Witten for illuminating correspondences
    on these points.}

\item
  In this work, however, we consider a different variant of the model,
  namely \eq{eq:Liou-cov-bdy}, which, in addition to the term in
  \eq{eq:polyakov} includes a negative cosmological constant and
  boundary terms, and consequently defines a theory of gravity in
  asymptotically AdS$_2$ spaces.

\item
  As we will find, the only physical degrees of freedom of
  \eq{eq:Liou-cov-bdy}, reduce to $M$, parametrized by $f(\tau)$ (see,
  e.g. \eq{non-norm-f}) which lives on the boundary. The bulk-boundary
  correspondence in this case essentially follows from two-dimensional
  diffeomorphism (this is somewhat reminiscent of Chern-Simon theories
  on a manifold with boundaries, or of AdS$_3$/CFT$_2$ duality). We
  will also find that the action describing the modes $f(\tau)$ is the
  Schwarzian action of SYK-type model and that the low temperature
  thermodynamics also have qualitative agreement with that of SYK.

\item
We would like to emphasize that while \eq{eq:polyakov}, in the gauge
\eq{alekseev-gauge}, arises from a coadjoint orbit action of
\emph{Diff}, we do not yet have an explicit proof that our proposed
bulk dual, described by \eq{eq:Liou-cov-bdy}, is also a coadjoint
orbit action of \emph{Diff} for asymptotically \ads geometries in
some gauge. While this may eventually turn out to be true, the
verification of our proposed duality in the rest of paper is
independent of such a connection.
  
\end{enumerate}

\section{Solutions of equations of motion and the Liouville action
\label{sec:Indu-grav-to-Liou}}

In this section we will first discuss the equations of motion from the
action \eq{eq:Liou-cov-bdy}. We will find that the solutions describe
spacetimes of constant negative curvature, which include \ads as well
as a three-parameter `non-normalizable' deformation, which correspond to geometries whose boundary is displaced with respect to the original boundary of \ads. We will
subsequently discuss the on-shell action.

\subsection{Equations of motion}

We now discuss the solutions of the above action,
\eqref{eq:Liou-cov-bdy}. We relegate the details of the computations
of the equations of motion to \autoref{sec:var-Poly} and summarize
only the important results here. The equations of motion are,
\begin{equation}\label{eq:eom-bulk-exact-text}\begin{aligned}
	0 &= \f1{16\pi b^2} \Bigg( g_{\mu\nu}(w) \Big(2 R(w) + 8 \pi\mu \Big) +  \int_\Ga ^x \bqty{ - 2 \cdev_\mu^{(w)} \cdev_\nu^{(w)} G(w,x) R(x) } \\
	& \hspace{15pt}+  \int_\Ga ^x \int_\Ga ^y  \bqty{ \partiald{G(w,x)}{w^\mu} \partiald{G(w,y)}{w^\mu} - \f12 g_{\mu\nu}(w) g^{\al\be}(w) \partiald{G(w,x)}{w^\al} \partiald{G(w,y)}{w^\be}} R(x) R(y) \Bigg)
\end{aligned} \end{equation}
It is more instructive to study the trace and traceless part of the equations separately,\footnote{We will subsequently write the action, \eqref{eq:Liou-cov-bdy} itself as sum over the trace and traceless part.}
\begin{align}
	\text{Trace part:} & \quad R(x) = - 8 \pi\mu  \label{eq:eom-trace} 
\\
	\text{Traceless part:} & \quad 0 =   \int_\Ga ^x \bqty{ - 2 \pqty{\cdev_\mu^{(w)} \cdev_\nu^{(w)} G(w,x) - \f12 g_{\mu\nu}(w) \Box^{(w)} G(w,x) } R(x) } \nonumber\\
	& \hspace{15pt}+  \int_\Ga ^x \int_\Ga ^y  \bqty{ \partiald{G(w,x)}{w^\mu} \partiald{G(w,y)}{w^\mu} - \f12 g_{\mu\nu}(w) g^{\al\be}(w) \partiald{G(w,x)}{w^\al} \partiald{G(w,y)}{w^\be}} R(x) R(y) \label{eq:eom-traceless}
\end{align}
Note that since $\mu>0$, the first equation, \eqref{eq:eom-trace},
signifies that the metric must have a constant negative curvature,
which of course includes \ads. Does \ads also satisfy
\eq{eq:eom-traceless}? What is the most general solution of both
equations?

We will leave details to \autoref{sec:var-Poly}, and state the main
results here. Let us write the metric in conformal gauge around an \ads
background, $g_{\al\be} = e^{2\phi} \gh_{\al\be}$, where
\begin{align}
\widehat{ds^2}  \equiv \gh_{\al\be} dx^\mu dx^\nu = \f1{\pi\mu (z+\zb)^2} 
\, dz\,d\zb = \f1{4\pi\mu \ze^2} \pqty{d\ze^2+d\tau^2}
\label{ads2}
\end{align}
Eq. \eqref{eq:eom-trace} then becomes the same as Liouville equation of
motion (see below for detail),
\begin{align}
	& \boxed{2\Boxh \phi = \Rh + 8\pi\mu e^{2\phi}}
\label{eq:eom}
\end{align}
which has the general solution \cite{Seiberg:1990eb,Witten:1976ck},
\begin{equation}\label{eq:sol-liou-mode-text}
	\phi = \f12 \log\qty[(z+\zb)^2 \frac{\del g(z) \delb \gb(\zb)}{(g(z)+\gb(\zb))^2}]
\end{equation}
where $g,\gb$ are arbitrary complex functions, conjugate of each other.\footnote{In Lorentzian signature, these functions can be chosen to be two independent real functions.} In the same gauge and background, \eqref{eq:eom-traceless} gives us the following \emph{Virasoro constraints} ,
\begin{equation}\label{virasoro-const}
\boxed{\begin{aligned}
\del^2\phi (z,\zb)- \pqty{\del\phi(z,\zb)}^2 + 2 \dfrac{\del\phi(z,\zb)}{z+\zb}
=0, \kern20pt 
\delb^2 \phi (z,\zb)- \pqty{\delb\phi(z,\zb)}^2 + 2 \dfrac{\delb\phi(z,\zb)}{z+\zb}=0
\end{aligned}
}
\end{equation}
Solving \eqref{eq:eom-trace} and \eqref{eq:eom-traceless} (or,
equivalently \eqref{virasoro-const}) simultaneously, 
we get solutions \eqref{eq:sol-liou-mode-text} 
with following conditions on $g,\gb$,
\begin{equation}\label{eq:Virasoro-constraints-text}\begin{aligned}
	\qty{ g(z), z} = 0, \quad \qty{ \gb(\zb), \zb} = 0 
	\Rightarrow g(z) = \frac{a z+ i b }  { i c z + d }, 
\quad \gb(\zb) = \frac{\bar a \zb - i \bar b }  {- i \bar c \zb + \bar d }, 
\quad a,b,c,d \in \mathbb{C}. \quad \footnotemark
\end{aligned}\end{equation}
\footnotetext{Here the independent set of parameters are constrained by $ad+bc=1$, which is the same as \slc.}
Here, and subsequently in this paper we denote the Schwarzian derivative of a function, $f(\tau)$, by $\qty{f(\tau),\tau} = \frac{f'''(\tau)}{f'(\tau)} - \f32 \pqty{\frac{f''(\tau)}{f'(\tau)}}^2$.
Of these solutions, the choice $a,b,c,d \in \real$ corresponds to \slr 
transformations of \ads coordinates, and are the exact isometries of 
the geometry. 

{\it The remaining 3-parameter set of solutions, which corresponds to
  the point marked NAdS$_2$ in Figure \ref{fig-coadjoint} are the
  solutions of our primary interest.}  These do not preserve the
boundary of \ads\!\!. In general, the boundary of the spacetime is given by the curve, $g(z)+\gb(\zb)=0$, which for a general function of the kind, \eqref{eq:Virasoro-constraints-text}, is not the same as $z+\zb=0$. These solutions will subsequently be referred to as
\emph{non-normalizable} solutions following the standard AdS/CFT language. 

The set of non-normalizable solutions obtained above is clearly
parameterized by $(a,b,c,d) \in $\slc/\slr, which can be identified
with a hyperboloid (see Appendix \ref{app:off-shell}, especially
\eq{hyperboloid} for more details).  The point $(a,b,c,d)=(1,0,0,1)$
corresponds to the identity transformation $g(z)=z$ in
\eq{eq:Virasoro-constraints-text}. We are interested in \emph{small}
non-normalizable deformations near the identity transformation. It is
possible to choose a set of coordinates of \slc/\slr, in which such
deformations are given by
\begin{equation}\label{eq:small-def-text}
	a = 1+ i\, \de a^I \quad
	b = i\,\de b^I \quad
	c = i\,\de c^I \quad
	d = 1 - i\, \de a^I \quad,
\end{equation}
where $\de a^I, \de b^I, \de c^I$ are real numbers. With these
parameters, the solution for the metric becomes
\begin{align}
ds^2  =  e^{2\phi} \widehat{ds^2}
\label{non-norm}
\end{align}
with $\widehat{ds^2}$ given by the \ads metric \eq{ads2}, and
$\phi$, using \eqref{eq:sol-liou-mode-text} and
\eqref{eq:Virasoro-constraints-text} has the near-boundary form
\begin{equation}\label{eq:de-g}
    \phi = -\frac{\de g(i\tau)}{\ze } + \mathcal{O}(\de a^2,\de b^2,
    \de c^2), \qquad -\de g(i \tau) = \de b^I + 2 \de a^I \tau +
    \de c^I \tau^2 
\end{equation}

Eventually, we will choose $\delta a^I= \delta c^I=0$, so that $\delta
g= - \delta b^I$, and $\phi = \de b^I/\ze$. We will find that the
$\delta b^I$ deformation (more precisely, $-\delta b^I$) corresponds
to the irrelevant coupling $1/J$ of the SYK model. The other
parameters $\delta a^I $ and $\delta c^I$ are physically distinct; it
would be interesting to explore their significance, which we leave for
future work.

For the Liouville factor $e^{2\phi}$ not to destroy the asymptotic
\ads structure altogether, we will assume here that $\delta g \lsim
\delta$; this ensures that $\delta g < \zeta$.\footnote{There is a
  natural RG interpretation of this inequality in terms of the
  boundary theory. We will later identify $\delta g$ with $\sim 1/J$
  (see \eq{parameter-match}). Together with the natural identification
  of $1/\ze$, for small $\ze$, with a Wilsonian floating cut-off
  $\Lambda$ (to be distinguished from the bare cut-off
  $\Lambda_0=1/\delta$, see \cite{Heemskerk:2010hk, Faulkner:2010jy},
  also \cite{Mandal:2016rmt}), we find $\delta g/\ze \sim \Lambda/J =
  1/\bar J$, where $\bar J= J/\Lambda$ is the dimensionless
  coupling. Since $\bar J$ grows large near the IR cut-off, it follows
  that $\delta g/\ze \ll 1$ near the IR cut-off.}
Note that the expression for the Liouville field in \eq{eq:de-g} is
similar to that of the dilaton in \cite{Maldacena:2016upp}, and plays
a somewhat similar role as we will see later.
In the next section, we will generate more solutions from the above
three-parameter solutions by using {\it large diffeomorphisms}, which
we cannot capture staying within the conformal gauge.  

\subsection{Liouville action}

We now show that the above analysis of equations of motion with
separation into trace and traceless parts also works for the classical
action. Writing the induced gravity action in a \emph{conformal gauge}
around an arbitrary fiducial metric, $\h g_{\al\be}$, we get the
action,\footnote{Later in this paper we will choose the fiducial
  metric from a class of Asymptotic \ads (A\ads) geometries. Although
  none of the analysis depends on the choice of this fiducial metric,
  it is only economical for a classical analysis that we choose it to
  be one of the saddle point solutions.}
\begin{equation}\label{eq:Scov-to-Liou}
    \begin{aligned}
	S_{cov}[g] &= -\f1{4\pi\; b^2} \Bigg[ \int_\Ga \sqrt{\gh} \qty( \gh^{\al\be} \del_\al \phi \del_\be \phi  + \Rh \phi + 4 \pi \mu e^{2\phi} ) +  2 \int_{\partial\Ga} \sqrt{\hat{\ga}} \hat{\Kb} \phi + \int_{\partial\Ga} \sqrt{\hat{\ga}} \hat{n}^\mu \phi \del_\mu \phi \\
	& \hspace{2cm}- \frac{1}{2} \int_{\partial\Ga} \sqrt{\hat{\ga}} \hat{n}^\mu \ \hat{\cdev}_\mu \pqty{ \phi \f1{\hat{\Box}}\Rh}\Bigg] + \f1{16\pi\; b^2}\int_\Ga \sqrt{\gh} \Rh \ \f 1{\h \Box} \Rh + \f 1 {4\pi\; b^2} \int_{\del\Ga} \sqrt{\ga} \h \Kb \ \f 1{\h \Box} \Rh \\[5pt]
		   &= -\f1{4\pi\; b^2} \Bigg[ \int_\Ga \sqrt{\gh} \qty( \gh^{\al\be} \del_\al \phi \del_\be \phi      + \Rh \phi + 4 \pi \mu e^{2\phi} ) +  2 \int_{\partial\Ga} \sqrt{\hat{\ga}} \hat{\Kb} \phi + \int_{\partial\Ga} \sqrt{\hat{\ga}} \hat{n}^\mu \phi \del_\mu \phi \Bigg]\\
		   & \hspace{2cm} + \f1{16\pi\; b^2}\int_\Ga \sqrt{\gh} \Rh \ \f 1{\h \Box} \Rh 
    \end{aligned}
\end{equation}
In all the above equations, the coordinate dependence of the functions
is understood. In the second line above, we have dropped the terms
boundary terms containing the Green's function, $\f1{\Boxh}$, given
the fall-off properties of the Green's function.  We
identify the part of the action in \eqref{eq:Scov-to-Liou} which
depends on $\phi$ field with Liouville action on a background with
metric $\h g$.
\begin{equation}\label{eq:S-Liou}
    S_L\qty[\phi,\h g] = -\f1{4\pi\, b^2} \Bigg[ \int_\Ga \sqrt{\gh} \qty( \gh^{\al\be} \del_\al \phi \del_\be \phi      + \Rh \phi + 4 \pi \mu e^{2\phi} ) +  2 \int_{\partial\Ga} \sqrt{\hat{\ga}} \hat{\Kb} \phi + \int_{\partial\Ga} \sqrt{\hat{\ga}} \hat{n}^\mu \phi \del_\mu \phi \Bigg]
\end{equation}
We are interested in computing the above action in the classical
limit, $b\to0$. The classical equation of motion for the $\phi$ field
turns out to be exactly the same as \eq{eq:eom}, the trace part of the equations of motion coming from the Polyakov action, as
expected. We emphasize the fact that if one chooses to study
\eqref{eq:Scov-to-Liou} as a theory of quantum gravity, then the trace
of background metric appearing there should not be treated as
independent degree of freedom.

One-dimensional Liouville equation of motion has appeared in
\cite{Engelsoy:2016xyb, Bagrets:2016cdf, Cotler:2016fpe} in the
context of boundary dynamics. However, their connection to the induced
gravity action that we have discussed here is not clear.

\paragraph{No dynamical Liouville mode:} It is 
important to note that in our problem 
there are no dynamical Liouville modes at all. 
The Liouville mode is entirely fixed in terms
of three parameters which, furthermore, correspond to non-normalizable
modes. These are specified as boundary conditions of the path integral
and are not dynamical variables. We elaborate on this point further in
Appendix \ref{app:off-shell} where we show that the form
of the Liouville mode, with three real constants, is completely
fixed by the Virasoro constraints alone.

\section{Asymptotically \texorpdfstring{\ads}{AdS2} geometries}\label{sec:asymp-ads}

In this section, we will construct asymptotically AdS$_2$ geometries
as a \emph{Diff} orbit of the solutions constructed in \eq{non-norm} ({\it
  see the orbits in the right panel of Figure \ref{fig-coadjoint}}).
To begin with, we will construct these asymptotic geometries purely
kinematically, from an analysis of asymptotic Killing vectors (AKV) of
\ads geometry (also see \autoref{app:asymp-ads-comp} for some details).  Later, we argue that
they solve the equations of motion and evaluate the on-shell action
for these configurations. AKV's of AdS$_2$ have been studied
earlier in \cite{Hotta:1998iq, Cadoni:1999ja} in the near-boundary
region, inspired by earlier work of Brown and Henneaux in one higher
dimension \cite{brown1986}.  We show below that it is possible to
integrate the infinitesimal diffeomorphisms exactly to find the full
nonlinear solution. This will lead to a class of A\ads geometries that
are related to each other by diffeomorphisms that become tangential at
the boundary. These geometries are dual to the conformally transformed
states in the 1-D field theory.\footnote{As indicated before,
  precisely at the conformal point, the stress tensor vanishes
  trivially; hence all states are ground states. However, slightly
  away from the conformal point, the (broken) conformal
  transformations lead to nontrivial states.}  We mainly consider
Euclidean metrics below.

Euclidean \ads metric in Poincare coordinates is defined by
\eq{ads2}.
The A\ads geometries are defined by the fall-off conditions \cite{Hotta:1998iq, Cadoni:1999ja, brown1986},
\begin{equation}\label{eq:fall-off}
    g_{\ze\ze} = \f1{4\pi\mu \, \ze^2} + \mathcal{O}(\ze^0), \quad g_{\ze\tau} = \mathcal{O}(\ze^0), \quad g_{\tau\tau} =  \f1{4\pi\mu \, \ze^2} + \mathcal{O}(\ze^0)
\end{equation}
Variation of the metric under most general diffeomorphism is,
\begin{align}\label{eq:metric-diff-var}
    \de g_{\al\be} = \cdev_\al \ep_\be + \cdev_\be \ep_\al = \begin{pmatrix} -\dfrac{\ep^\ze (\ze,\tau)-\ze \del_\ze\ep^\ze(\ze,\tau)}{2 \pi \mu \ze ^3} & \dfrac{\del_\tau \ep^\ze(\ze,\tau)+\del_\ze\ep^\tau(\ze,\tau)}{4 \pi \mu \ze^2 }\\[-1pt]  \dfrac{\del_\tau \ep^\ze(\ze,\tau)+\del_\ze\ep^\tau(\ze,\tau)}{4 \pi \mu \ze^2 } & -\dfrac{\ep^\ze(\ze,\tau)-\ze \del_\tau\ep^\tau(\ze,\tau)}{2 \pi \mu  \ze^3 }   \end{pmatrix}
\end{align}
The asymptotic Killing vectors can be solved for by imposing on \eqref{eq:metric-diff-var} the fall-off conditions in \eqref{eq:fall-off}, \cite{Cadoni:1999ja,Hotta:1998iq}. However, we choose to work in Fefferman-Graham gauge which is defined by,
\begin{equation}\label{eq:FG-gauge-cond}
    \de g_{\ze\ze} = 0, \quad \de g_{\ze\tau} = 0
\end{equation}
The solution for the asymptotic Killing vectors is given in terms of an arbitrary function, $\de f(\tau)$,
\begin{equation}\label{eq:asymp-kill-vec}
    \ep^\ze(\ze,\tau) = \ze \de f'(\tau), \quad \ep^\tau(\ze,\tau) = \de f(\tau) - \f12 \ze^2 \de f''(\tau)
\end{equation}
It is clear from the above solution, that the diffeomorphism is tangential at the boundary of \ads, $\ze=0$. The integrated form of the coordinate transformations is,
\begin{align}\label{eq:exact-large-diff}
\boxed{\taut = f(\tau) - \frac{2 \ze^2 f''(\tau) f'(\tau)^2}{4 f'(\tau)^2 + \ze^2 f''(\tau)^2}, \quad \zet = \frac{4 \ze f'(\tau)^3}{4 f'(\tau)^2 + \ze^2 f''(\tau)^2}}
\end{align}
Although we think that this choice of gauge should not be necessary and it should be possible to integrate the diffeomorphisms more generally, we found it easier to do so with this gauge choice. This was largely motivated by \cite{Banados:1998gg, Roberts:2012aq} who performed similar integrations of diffeomorphisms in AdS$_3$ case. The details of this computation are presented in \autoref{app:asymp-ads-comp}.

The result of this diffeomorphism can be stated as follows. If we start
with the \ads metric in the $\zet$-$\taut$ coordinates
\begin{align}
\widehat{ds^2} = \f1{4\pi\mu\, \zet^2} \pqty{ d\zet^2 + d\taut^2}, \nonumber
\end{align}
in the original $\ze$-$\tau$ coordinates it becomes
\begin{align} 
\boxed{\widehat{ds^2} = \f1{4\pi\mu\, \ze^2} \pqty{ d\ze^2 +  d\tau^2 \pqty{1- \ze^2 \frac{\qty{f(\tau),\tau}}{2}}^2 }} \label{eq:AAds-class}
\end{align}
Recall that $\qty{f(\tau),\tau} = \frac{f'''(\tau)}{f'(\tau)} - \f32 \pqty{\frac{f''(\tau)}{f'(\tau)}}^2$ is the standard notation for Schwarzian derivative that we use throughout this paper. We want to emphasize that the class of geometries given by \eqref{eq:AAds-class} also have constant negative curvature, $\h R = - 8 \pi \mu$. As in AdS$_3$, it should be possible to identify these geometries as different sections of the global \ads geometry. Some discussion of how various \ads geometries are related is provided in \cite{Cadoni:1999ja}.

One can carry out the above diffeomorphism in the presence of the
non-normalizable solutions described in the previous section.
To do this, we 
begin with the metric \eq{non-norm} in the $\ti \ze$-$\ti \tau$
coordinates:
\begin{align}
ds^2 =  e^{2\tilde \phi(\tilde x^\mu)} \widehat{ds^2}, \kern10pt
\tilde \phi(\tilde x^\mu) &=\f{\tilde {\de g}}{\ti \ze} +
\mathcal{O}(\de a^2,\de b^2,
    \de c^2), \; 
\tilde {\de g}= \Im(\de b) + 2 \Im(\de a) \ti \tau +
\Im(\de c){\ti \tau}^2
\nonumber
\end{align}
and transform to $\ze$-$\tau$ coordinates, yielding
the metric
\begin{empheq}[box=\fbox]{align}
\label{non-norm-f}
ds^2 &=  e^{2\phi} \widehat{ds^2}, \, \widehat{ds^2}=
\f1{4\pi\mu\, \ze^2} \pqty{ d\ze^2 +  
d\tau^2 \pqty{1- \ze^2 \frac{\qty{f(\tau),\tau}}{2}}^2 }, 
\nonumber\\
\phi &=-\f{\de g(i\taut)}{\zet(\ze,\tau)} + \mathcal{O}(\de a^2,\de b^2,
    \de c^2), \; 
-\de g(i \taut) = \de b^I + 2 \de a^I \taut +
    \de c^I \taut^2,\, \taut= f(\tau)
\end{empheq}  

{\it In terms of the Figure \ref{fig-coadjoint}, the above solutions
  \eq{eq:AAds-class}, \eq{non-norm-f} represent the Diff orbit of \ads
  and N\ads on the right panel.}

As remarked below \eq{eq:de-g}, we eventually choose only the
one-parameter deformation parameterized by $\delta b^I$, which will
turn out to correspond to the $1/J$ deformation of the strong coupling
fixed point of the SYK theory. However, for the sake of generality,
we will for now continue with the more general form of $\delta g$.

\subsection{Proper treatment of the bulk path integral}\label{subsec:proper-path}
To this point we have not discussed the issue of gauge fixing inside the quantum mechanical path integral. While we are largely interested in a classical computation in the bulk, where the path integral measure due to gauge fixing is not important, we now shed some light on this issue. The computation of the ghost action is discussed in detail in \autoref{sec:quantum-corr}. The gauge fixing $\de$-function and the corresponding Faddeev-Popov determinant is given by,
\begin{equation}\label{eq:delta-def-text}
	1 = \De_{FP}\Big[ \gh[f(\tau)],\phi \Big] \times \int [\cD \ep^{(s)}] [\cD \phi] [\cD f(\tau)] \; \de\qty( g^{\e^{(s)}} - e^{2\phi} \gh[f(\tau)] ) \times \de\qty(\ep^{(s)}(z_1)) \, \de\qty(\ep^{(s)}(z_2)) \, \de\qty(\ep^{(s)}(z_3))
\end{equation}
In line with the discussion of the previous sections, we gauge fix an arbitrary metric to be conformally related to the A\ads metrics. In the choice of this gauge, there is an additional \slr residual gauge freedom that has been fixed using the $\de$-functions that anchor three arbitrary points in the geometry.\footnote{This is the standard prescription followed in open-string path integral computations. See also the relevant discussion in \cite{Maldacena:2016hyu}.} Going through the standard procedure of introducing the fermionic ghosts, we obtain a ghost action \eqref{eq:FP-det}. This procedure should not only capture the correct Jacobian required for the gauge fixing, but also for defining an invariant measure on the space of $f(\tau)$ integrations. 

With the above ingredients, the path integral is given by
\begin{align}
Z=  \int \frac{{\cD f(\tau)}'}{f'(\tau)} 
\exp[-S_{hydro} + ...]
\label{hydro-z}
\end{align}
where $S_{hydro}$ is the effective action \eq{schwarzian-summary},
describing the hydrodynamic modes (see the next section). The terms in
the ellipsis denote subleading terms which get contribution from the
Faddeev-Popov determinant mentioned above and discussed in detail in
\autoref{sec:quantum-corr}. The integration measure is the invariant
integration measure in the space of $f(\tau)$ functions.
The \emph{prime} on the measure denotes the exclusion of the
integration over \slr~degrees of freedom due to the treatment of
\slr~modes discussed above.\footnote{This measure should
appear from a proper treatment of the Faddeev-Popov
procedure which is sketched in \autoref{sec:quantum-corr}. We leave
details of this to  subsequent work.}

\section{Action of hydrodynamics modes}\label{sec:Hydrodynamics}
We now compute the on-shell action of the above geometries to
determine the contribution of the {\it large diffeomorphisms}
discussed in this section to the partition function of the system.

	\subsection{Boundary action}\label{subsec:bdy-act}
	We know from our analysis of equations of motion in \autoref{sec:var-Poly} that all of A\ads geometries satisfy the bulk equations of motion. Thus we can safely anticipate that the major contribution to the action of hydrodynamics modes will come from the boundary terms in \eqref{eq:Scov-to-Liou}. The boundary terms of the action are given by,
	\begin{equation}\label{eq:just-bdy}
    		S^{bdy}_L\qty[\phi,\h g] = -\f1{4\pi\, b^2} \Bigg[  2 \int_{\partial\Ga} \sqrt{\hat{\ga}} \hat{\Kb} \phi + \int_{\partial\Ga} \sqrt{\hat{\ga}} \hat{n}^\mu \phi \del_\mu \phi \Bigg]
	\end{equation}
	The second term above doesn't contribute at the leading order. The contribution of this term starts at, $\mathcal{O}(\de g)^2$ and hence won't contribute to the leading order answers that we subsequently compute.\\
	We also emphasize on the correct way to regulate the geometries for the subsequent computations. To keep our notations unambiguous, we will denote the coordinates of \ads by $\zet, \taut$ and that of A\ads geometries by $\ze,\tau$. We know that A\ads geometries are related to \ads geometry by large diffeomorphisms. If we put a radial cut-off in \ads at $\zet=\de$ and apply these large diffeomorphisms, then the cut-off at constant $\zet$ is mapped to some \emph{wiggly-curves} in $\ze$-$\tau$ coordinates,\footnote{There are two solutions for $\ze$ satisfying $\zet=\de$ (because the second equation in \eqref{eq:exact-large-diff} is a quadratic in $\ze$), one of which doesn't satisfy the boundary condition, $\ze\to0$ as $\de\to0$.}
	\begin{equation}\label{eq:mod-bdy-curve}\begin{aligned}
		&\de = \frac{4 \ze f'(\tau)^3}{4 f'(\tau)^2 + \ze^2 f''(\tau)^2}\\[5pt]
		\Rightarrow & ~\ze = \frac{2 \, f'(\tau)}{\de \, f''(\tau)^2} \left[ f'(\tau)^2 - \sqrt{f'(\tau)^4 - \de^2 \, f''(\tau)^2} \right]
	\end{aligned}\end{equation}
	These are the same wiggles as discussed in \cite{Maldacena:2016upp}. To consider physically distinct geometries in $\ze$-$\tau$ coordinates, we put a cut-off at $\ze=\de$ and compare the action with that of geometry corresponding to $\zet=\de$.

	\paragraph{In \ads} On the boundary $\zet=\de$, $\sqrt{\h\ga} \h \Kb = \f1\de$
	\begin{align}
		S^{bdy}_L[\phit, \ti g_{\al\be}] =-\f1{2\pi\, b^2} \int_{\partial\Ga} \sqrt{\hat{\ga}} \hat{\Kb} \phit =- \f1{2\pi\, b^2 } \int d\taut \bqty{ \pqty{ \frac{\de g(i \taut)}{\de^2} - \frac{1}{2} \ \de g''(i \taut) + \mathcal{O}(\de^2) } + \mathcal{O}\qty[\de g(i\taut)^2] }
	\end{align}
	where, $\de g(\taut)$ was defined in \eqref{eq:de-g}. To be able to compare with the A\ads answer later, we do the coordinate transformation from $\taut \to \tau$ coordinates,
	\begin{align}\label{eq:ads-bdy-term}
	    S^{bdy}_L[\phit, \ti g_{\al\be}] =-\f1{2\pi\, b^2} \int_{\partial\Ga} \sqrt{\hat{\ga}} \hat{\Kb} \phit &= -\f1{2\pi\, b^2 \; \de} \int d\tau \partiald{\taut(\tau)}{\tau} \Bigg[ \pqty{ \frac{\de g(i \taut(\tau))}{\de} - \frac{\de}{2} \ \de g''(i \taut(\tau)) + \mathcal{O}(\de^2) } \nonumber \\
		&\hspace{3cm}+ \mathcal{O}\qty[\de g(i\taut(\tau))^2]\Bigg] \\
		\taut &= f(\tau) - \frac{f'(\tau)^2}{f''(\tau)} \pqty{1 - \sqrt{1- \de^2 \pqty{ \frac{f''(\tau)}{f'(\tau)^2} }^2} } 
	\end{align}
	Here, it is important to note that we need to implement the coordinate transformation at the $\zet=\de$ slice. To this effect, we need to solve for $\ze$ at $\zet=\de$ using the second equation in \eqref{eq:exact-large-diff} and substitute it back in the first equation there.

	\paragraph{In A\ads} On the boundary $\ze=\de$, $\sqrt{\h\ga} \h \Kb = \f1\de + \de \frac{\qty{f(\tau),\tau}}{2}$
	\begin{align}\label{eq:aads-bdy-term}
	    S^{bdy}_L[\phi, g_{\al\be}] = -\f1{2\pi\, b^2} \int_{\partial\Ga} \sqrt{\hat{\ga}} \hat{\Kb} \phi &=- \f1{2\pi\, b^2 } \int d\tau \bqty{ \f1\de + \de \frac{\qty{f(\tau),\tau}}{2} } \times \Bigg[ \Bigg( \f1\de \ \frac{\de g(i f(\tau))}{ f'(\tau)} \nonumber \\
		&\hspace{-20pt}- \de \ \frac{ \Big(-\text{$\delta $g}(i f(\tau)) f''(\tau)^2+2 f'(\tau)^4 \text{$\delta $g}''(i f(\tau))+2 i f'(\tau)^2 f''(\tau) \text{$\delta $g}'(i f(\tau))\Big)}{4 f'(\tau)^3} \nonumber\\
	& + \mathcal{O}(\de^2)  \Bigg) + \mathcal{O}\qty[\de g(if(\tau))^2] \Bigg]
	\end{align}

	Hence,
	\begin{equation}\begin{aligned}\label{eq:bdy-diff}
		\de S^{bdy}_L  = S^{bdy}_L[\phi,g_{\al\be}] - S^{bdy}_L[\phit,\ti g_{\al\be}] &=\f1{2\pi\, b^2} \int d\tau\Bigg[ \bigg(\frac{\de g(i f(\tau ))}{\de^2} \left(f'(\tau )-\frac{1}{f'(\tau)} \right) - \frac{\de g(i f(\tau ))}{f'(\tau )} \qty{f(\tau),\tau} \\
		&\hspace{3cm}+ \mathcal{O}(\de^2) \bigg) + \mathcal{O}\bqty{\de g(i f(\tau))^2} \Bigg]\\     
	\end{aligned}\end{equation}
The $O(1/\de^2)$ divergent term can be  subtracted by introducing following counterterm in the action \eqref{eq:S-Liou},
\begin{align}\label{eq:counterterm}
S_{ct}=  \f{4\sqrt{\pi\mu}}{4\pi\, b^2} \int_{\partial\Ga} \sqrt{\hat{\ga}} \phi 
\end{align}
which essentially replaces $\sqrt{\hat\ga}\hat{\Kb} \to \sqrt{\hat\ga} (\hat{\Kb} -1)$.\footnote{A
  similar counterterm is also implied in \cite{Maldacena:2016hyu} in
  removing a quadratic divergence from their computation of the
  Schwarzian term.} To the linear
order in $\delta g$ under consideration here, this is the same as the
fully covariant counterterm $-4\sqrt{\pi\mu} \f1{4\pi b^2} \int_{\del\Ga}
\sqrt{\ga}\,\f1{\Box} R$. The finite part of the answer is,

	\begin{equation}\begin{aligned}\label{hydro-first}
	 \de S^{bdy}_L = -\f1{2\pi b^2} \int d\tau \frac{\de g(i
           f(\tau ))}{f'(\tau )} \qty{f(\tau),\tau} = \f1{2\pi b^2}
         \int d\taut\ \de g(i \taut) \; \qty{\ft(\taut),\taut }
	\end{aligned}\end{equation}
Here the third term is written in terms of the $\taut$ coordinate, the
boundary coordinate of the unperturbed \ads\!\!.\footnote{In going
  from the second expression to the third term, we have first
  transformed to the time coordinate $\taut=f(\tau)$, with $\tau=
  \tilde f(\taut)$, $\tilde f\equiv f^{-1}$, and used the Schwarzian
  composition rule
\[
\{\tilde f(f(\tau)), \tau\} = 
\{\tilde f(f(\tau)), f(\tau)\} f'(\tau)^2 +  \{f(\tau), \tau\}
= \{\tilde f(\taut), \taut\} f'(\tau)^2 +  \{f(\tau), \tau\},
\]
The LHS equals $\{\tau, \tau\}$ and vanishes.} Also, note that we have defined $\tilde f(\taut) = \tau$ as the reparametrized coordinate starting with the unperturbed \ads\!\! coordinate $\taut$.\footnote{It is important to note that the large diffeomorphism
  $\tilde f$ is what corresponds to the pseudo-Nambu-Goldstone mode $f$
  of the SYK model.}

The function $\delta g(\tau)$ is given by \eq{eq:de-g}.  As indicated
below that equation, we will henceforth choose $\delta g$ = constant.
One might wonder if one can absorb the $\tau$ and $\tau^2$
deformations in $\delta g$, parameterized by $\delta a^I$ and $\delta
c^I$, by a possible reparameterization of the boundary
coordinate $\tau$; this, however, turns out impossible for any value
of these parameters since the corresponding transformation turns out to
be singular. Thus, the $\delta b^I, \delta a^I, \delta c^I$ represent
different physics, and we will find that it is only the $\delta b^I$
deformation, that is, a constant $\delta g$, which will correspond to the SYK
model. We will see that the non-normalizable mode
corresponding to constant $\delta g$, will correspond to the irrelevant
coupling $1/J$ of the SYK model. As remarked before, it is an
important open question what the other parameters $\delta a^I$ and
$\delta c^I$ correspond to. 

In \autoref{sec:matching} we will do the detailed matching with the
boundary field theory. Note that the \slr transformations that
correspond to the `global conformal transformations' of one
dimensional space remain the symmetry of this action. We have
presented a discussion of the correct measure of integration over the
$\ft(\tau)$ modes in \autoref{subsec:proper-path} and
\autoref{sec:quantum-corr}.

	\subsection{Bulk action}\label{subsec:bulk-act}
	The bulk part of the Liouville action is,
	\begin{align}
		S^{bulk}_L\qty[\phi,\h g] &= -\f1{4\pi\, b^2} \int_\Ga \sqrt{\gh} \qty( \gh^{\al\be} \del_\al \phi \del_\be \phi      + \Rh \phi + 4 \pi \mu e^{2\phi} ) \nonumber \\
		& = -\f1{4\pi b^2} \int_{\del\Ga} \sqrt{\h\ga} \h n^\al \phi \del_\al \phi - \f1{4\pi b^2} \int_{\Ga} \sqrt{\h g} \qty(-\phi \Boxh \phi + \Rh \phi + 4 \pi \mu e^{2\phi}) \nonumber \\
		& = -\f1{4\pi b^2} \int_{\del\Ga} \sqrt{\h\ga} \h n^\al \phi \del_\al \phi - \f1{4\pi b^2} \int_{\Ga} \sqrt{\h g} \qty(\f12 \Rh \phi + 4 \pi \mu e^{2\phi} \pqty{1-\phi}) \nonumber \\
		& = -\f1{4\pi b^2} \int_{\del\Ga} \sqrt{\h\ga} \h n^\al \phi \del_\al \phi - \f\mu{b^2} \int_{\Ga} \sqrt{\h g} \qty(-\phi + e^{2\phi} \pqty{1-\phi})
	\end{align}
	here in the second line we have shifted the derivatives, while in the second term we have used the equation of motion of the $\phi$ field, $\Boxh \phi = \f12 \Rh + 4\pi \mu e^{2\phi}$. In the last line we have substituted the value of $\Rh=-8\pi\mu$. The first term in the above equation contributes at subleading order, as argued under \eqref{eq:just-bdy}. Using the on-shell value of the $\phi$ we evaluate the above action in \ads and A\ads backgrounds.

\paragraph{In \ads}
	For \ads background metric, the action is given by,
	\begin{align}\label{eq:ads-bulk}
	    S_L^{bulk}[\phit, \ti g_{\al\be}] &= - \f\mu{b^2} \int\limits_{-\infty}^{\infty} d\taut \int\limits_{\zet=\de}^\infty d\zet \sqrt{\ti g} \qty(-\phit + e^{2\phit} \pqty{1-\phit}) \nonumber \\
	    &= - \f\mu{b^2}  \int\limits_{-\infty}^{\infty} d\tau \int\limits_{\ze>\del\Ga}^\infty d\ze \sqrt{g} \qty(-\phi + e^{2\phi} \pqty{1-\phi})
	\end{align}
	Here, in the second line we have used the coordinate transformations, \eqref{eq:exact-large-diff} and the boundary in $\ze$ coordinates is now given by the wiggly curve, \eqref{eq:mod-bdy-curve},
	\[
	    \del\Ga \equiv \ze =\f2{\de \; f''(\tau)^2} \bqty{f'(\tau)^3 - \sqrt{f'(\tau)^6 - \de^2 \, f'(\tau)^2 \, f''(\tau)^2}}
	\]

\begin{figure}[H]
		\begin{center}
			\includegraphics[width=8cm]{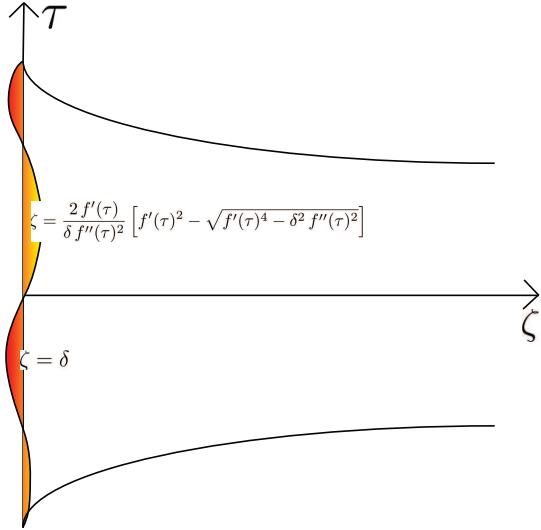}
		\caption{\footnotesize The difference in bulk action between \ads and A\ads geometries gets contribution only from the shaded region}
		\label{fig:diff-act}
		\end{center}
	\end{figure}

\paragraph{in A\ads} Similarly for the A\ads background we have the action,
	\begin{align}\label{eq:aads-bulk}
		S_L^{bulk}[\phi,g_{\al\be}] = -\f\mu{b^2} \int\limits_{-\infty}^\infty d\tau \int \limits_{\ze=\de}^\infty d\ze \sqrt{g} \pqty{ -\phi + e^{2\phi} (1-\phi) }
	\end{align}
	Thus we have,
	\begin{align}
		\de S_L^{bulk} = S_L^{bulk}[\phi,g_{\al\be}] - S_L^{bulk}[\phit, \ti g_{\al\be}] = -\f\mu{b^2} \int\limits_{-\infty}^\infty d\tau \int\limits_\de^{\ze=\del\Ga} d\ze \sqrt{g} \pqty{ -\phi + e^{2\phi} (1-\phi) }
	\end{align}
	It is easy to approximate this expression close to the boundary of the geometry, \emph{i.e.} when $\de\to0$. In that case the difference between $\ze=\del\Ga$ and $\ze=\de$ reduces to a small strip as shown in \autoref{fig:diff-act}.
	Moreover $(-\phi + e^{2\phi} (1-\phi)) \sim 1 + \mathcal{O}[\de^3, \de g(i \tau)^2]$ so that part of the integrand becomes trivial. Hence we can approximate the integrand by,
	\begin{equation}\label{eq:bulk-approx}
		\de S^{bulk}_L = \f{1}{4\pi b^2} \int\limits_{-\infty}^\infty d\taut \bqty{ \f1\de \pqty{\ft'(\taut)-1} + \f\de4 \pqty{ \frac{\left(2\ft'(\taut)-3\right) \ft''(\taut)^2-2 \ft'''(\taut) \left(\ft'(\taut)-1\right) \ft'(\taut) }{\ft'(\tau)^3} } }
	\end{equation}
The first term is linearly divergent,\footnote{We thank Shiraz
  Minwalla for a crucial discussion on this point.} however while considering the coordinate transformations which approach identity transformations asymptotically this term integrates to zero. In other words if we consider a transformation, $\ft(\taut) = \taut + \ep(\taut)$\footnote{Here $\ep(\taut)$ is not necessarily small, but just a rewriting of the coordinate transformations.}, then $\ft'(\taut) = 1 + \ep'(\taut)$. In this case the first term becomes,
  \begin{equation}
    \f{1}{4\pi b^2} \int\limits_{-\infty}^\infty d\taut \, \f1\de \ep'(\taut) = \f{1}{4\pi b^2 \, \de} \bqty{\ep(\infty) - \ep(-\infty)}
  \end{equation}
  Clearly a \emph{good} coordinate transformation has to be monotonically increasing. Additionally we require, for the transformation to remain invertible, that $\ep(\infty) = 0 = \ep(-\infty)$. In fact, the transformation that we use to map the theory on a line to a theory on a thermal circle is not of this kind and the regulation scheme adopted in that case is explained in \autoref{sec:Thermodynamics}. Leaving aside the issue of the regulation, the reparametrization of a thermal quantum mechanical theory can be achieved starting from a quantum mechanical theory in two steps: firstly, the straight line is mapped to a thermal circle using the map $\taut = \tan(\pi \thet/\be)$; then, one reparametrizes the thermal circle with appropriate boundary conditions, ensuring reparametrization doesn't change the winding around the circle and is invertible. In this case, the Schwarzian action becomes \cite{Maldacena:2016hyu,Maldacena:2016upp}:
  \begin{align}
  	S_{hydro}^\be=\frac{\de g}{2\pi b^2}\int d\thet \qty{ \frac{\be}{ 2 } \tan(\pi \f{f(\thet)}{\be}) ,\thet}
  \end{align}

\subsection{Summary}

We thus find that the following low energy effective action
(in the leading large $1/b$ limit) for the `hydrodynamic modes'
\begin{align}
\boxed{S_{hydro}  = \f{\delta g}{2\pi b^2} \int
d\taut\ \qty{\ft(\taut),\taut }}
\label{schwarzian-summary}
\end{align}
In \autoref{sec:matching} we will compare this with the Schwarzian
term which appears in the SYK-type models.

It is important to mention that the bulk dual discussed in
\cite{Almheiri:2014cka, Maldacena:2016upp}, leads to a similar
Schwarzian term starting from a dilaton gravity model, while the bulk
dual discussed in this paper has only the metric field described by
the Polyakov action. The source of the hydrodynamic modes in both
cases involves the {\it large diffeomorphisms} which are nontrivial at
the boundary. In a very recent paper \cite{Nakayama:2017pys}, another
proposal for a bulk dual has appeared which has a Liouville field {\it
  and} the Almheiri-Polchinski action \cite{Almheiri:2014cka}. They
also appear to get a Schwarzian term rather differently, from the
Liouville fluctuations similar to our functions $g(z), \gb(\zb)$ in
\eqref{eq:sol-liou-mode-text}. However, as we found above, except for
an \slr worth of degrees of freedom (see
\eq{eq:Virasoro-constraints-text}, \eq{eq:small-def-text}), these
Liouville fluctuations are frozen by the Virasoro gauge conditions
\eq{virasoro-const}. It is also pertinent here to mention the theorems
due to Schwarz and Pick \cite{Osserman:1998}; these restrict the class
of conformal transformations that map the boundary of Poincare
half-plane to itself to only \slr transformations.

\section{Thermodynamic partition function from bulk dual}\label{sec:Thermodynamics}
In this section we compute the Euclidean bulk partition function in the classical limit for a black hole geometry. We use the standard prescription of \cite{hawking1982} to renormalize the bulk partition function by subtracting the partition function of thermal \ads geometry from the Euclidean black-hole geometries that we describe below.\footnote{Thermal \ads geometry is obtained simply by identifying the boundary time coordinate in \eqref{ads2} over a period $\be$.} Following \cite{Maldacena:2016hyu, Maldacena:2016upp}, we can do a reparametrization of the Euclidean time to study a field theory defined on a thermal circle of length $\be$,
\begin{equation}\label{eq:thermal-repara}
	\taut = \tan(\frac{\pi \thet}{\beta})
\end{equation}
Using \eqref{eq:AAds-class}, we can compute the Euclidean geometry that is dual to the thermal field theory,
\begin{equation}\label{eq:thermal-bulk-Euc}
	ds^2 = \frac{1}{4 \pi \mu \, \zet^2} \bqty{ d\zet^2 + \pqty{1-\pi^2 \frac{\zet^2}{\be^2}}^2 d\taut^2 }, \quad \taut\in\qty(-\f\be 2,\f\be 2) \text{ and } \zet\in \qty(0,\f\be \pi)
\end{equation}
This geometry is a \emph{capped \ads geometry} in two dimensions. There is no deficit angle near the horizon of the geometry, which can be easily checked by doing a near horizon expansion, $\zet = \be/\pi - \rho$, $$ds^2 \sim \f\pi{4\mu \, \be^2} \bqty{ d\rho^2 + 4 \f{\pi^2}{\be^2} \rho^2 d\thet^2 }$$ Analytically continuing this geometry to Lorentzian space we get,
\begin{equation}\label{eq:thermal-bulk-Lor}
	ds^2 = \frac{1}{4 \pi \mu \, \zet^2} \bqty{ d\zet^2 - \pqty{1-\pi^2 \frac{\zet^2}{\be^2}}^2 dt^2 }
\end{equation}
which is a geometry with a horizon at $\zet=\be/\pi$. 

To get the free energy of the theory, we compute the on-shell bulk action for this geometry, but with a small non-normalizable deformation turned on (\emph{smallness} is understood as explained in the previous section).
\paragraph{Bulk action}
We first compute the bulk part of the action given in \eqref{eq:S-Liou}. The bulk part of the Liouville action is,
	\begin{align}
		S^{bulk}_L\qty[\phi,\h g] &= -\f1{4\pi\, b^2} \int_\Ga \sqrt{\gh} \qty( \gh^{\al\be} \del_\al \phi \del_\be \phi      + \Rh \phi + 4 \pi \mu e^{2\phi} ) \nonumber \\
		& = -\f1{4\pi b^2} \int_{\del\Ga} \sqrt{\h\ga} \h n^\al \phi \del_\al \phi - \f1{4\pi b^2} \int_{\Ga} \sqrt{\h g} \qty(-\phi \Boxh \phi + \Rh \phi + 4 \pi \mu e^{2\phi}) \nonumber \\
		& = -\f1{4\pi b^2} \int_{\del\Ga} \sqrt{\h\ga} \h n^\al \phi \del_\al \phi - \f1{4\pi b^2} \int_{\Ga} \sqrt{\h g} \qty(\f12 \Rh \phi + 4 \pi \mu e^{2\phi} \pqty{1-\phi}) \nonumber \\
		& = -\f1{4\pi b^2} \int_{\del\Ga} \sqrt{\h\ga} \h n^\al \phi \del_\al \phi - \f\mu{b^2} \int_{\Ga} \sqrt{\h g} \qty(-\phi + e^{2\phi} \pqty{1-\phi})
	\end{align}
	here in the second line we have shifted the derivatives, while in the third line we have used the equation of motion of $\phi$ field, $\Boxh \phi = \f12 \Rh + 4\pi \mu e^{2\phi}$. In the last line we have substituted the value of $\Rh=-8\pi\mu$. The first boundary term in the last line combines with the boundary term already present in \eqref{eq:S-Liou}. However, these terms are not important for our analysis because they only contribute at $\mathcal{O}(\de g^3)$. We don't have any leading contribution coming from the $\de g$ modes from the bulk action,
	\begin{equation}\label{eq:bulk-thermal}
	    S^{bulk}_L[\phit,\tilde g_{\al\be}] = \frac1{2b^2} - \frac{\be}{4\pi b^2 \, \de} - \frac{\pi \, \de}{4b^2 \, \be} + \mathcal{O}(\de g^3)
	\end{equation}
	As was the case with the previous Hydrodynamics calculation, all the divergent as well as finite terms above are cancelled by subtraction of the thermal \ads partition function. This is the standard prescription to regulate the partition function of the black hole geometries (see \cite{hawking1982}). Thus the bulk contribution starts only at $ \mathcal{O}(\de g^3)$.
	\paragraph{Boundary action} Computing the boundary terms of the action \eqref{eq:S-Liou}. Again, as argued above, the last term in \eqref{eq:S-Liou} doesn't contribute at leading order. The term containing extrinsic curvature when evaluated on the boundary gives,
	\begin{equation}\label{eq:bdy-thermal}
	    S^{bdy}_L[\phit,\tilde g_{\al\be}] = \frac{\de g}{2b^2 \, \be} + \frac{\be \, \de g}{4\pi^2 b^2 \, \de^2} + \mathcal{O}(\de g^2)
	\end{equation}
	In both the above expressions we have taken the boundary value of the $\de g(i \taut)$ field to be constant, as explained earlier, and have denoted it by $\de g$. Again, the quadratically divergent term is cancelled by inclusion of the counterterm discussed in \eq{eq:counterterm}.

    	One last piece that needs to be evaluated is the bulk term $\int\sqrt{\h g}\Rh \f1\Boxh \Rh$ that depends only on the background geometry. The Green's function in hyperbolic spaces is a well studied subject. In Green's function can be evaluated by taking a limit of the `resolvent' of the Laplacian.\footnote{A resolvent in defined as the classical Green's function of the operator $-\Box + 4\pi\mu s (s-1)$. Thus the required Green's function is the $s\to1$ limit of the resolvent.} The resolvent of the Laplacian on right half Poincare-plane, $\rhp$, is given by,
	\begin{align}\label{eq:resolvent}
	    \pqty{-\Boxh_{z} + 4 \pi \mu s(s-1) } R_{\rhp}(s; z,w ) = 4 \pi \mu \de^{(2)}(z-w) \nonumber \\
	    R_{\rhp}(s; z,w) = \f1{4\pi} \f{\Ga(s)^2}{\Ga(2s)} \pqty{1+\frac{|z-w|^2}{4 \Re(z)\Re(w)}}^{-s} \; {}_2F_1\bqty{s,s;2s;\frac1{1+\frac{|z-w|^2}{4 \Re(z)\Re(w)} } }
	\end{align}
	Here, $z,w$ are the complexified coordinates, $z=\ze_1+i \tau_1$ and $w=\ze_2+i \tau_2$. The $s\to1$ limit of this function is,
	\begin{equation}\begin{aligned}
		G\pqty{\{\ze_1,\tau_1\};\{\ze_2,\tau_2\}} = -\f1{4\pi} \log \left(1-\frac{4 \ze_1 \ze_2}{(\ze_1+ \ze_2)^2 + (\tau_1-\tau_2)^2}\right)
	    \end{aligned}\end{equation}

	However, the above results are in $\rhp$, while we are interested in solving the Green's function for the geometry in \eqref{eq:thermal-bulk-Euc}. The Green's function can be obtained easily using the coordinate transformations in \eqref{eq:exact-large-diff} with the choice of function in \eqref{eq:thermal-repara}. We get,
	\begin{equation}\label{eq:greens}\begin{aligned}
		G = -\f1{4\pi} \log\bqty{1-\frac{8 \pi ^2 \be ^2 \ze_1 \ze_2}{\be ^4+\pi ^2 \be ^2 \left(\ze_1^2+4 \ze_1 \ze_2+\ze_2^2\right)-(\be^2 -\pi^2  \ze_1^2) (\be^2 -\pi^2 \ze_2^2) \cos(\frac{2 \pi  (\theta_1-\theta_2)}{\be})+\pi ^4 \ze_1^2 \ze_2^2} }
	    \end{aligned}\end{equation}
	With this Green's function we solve the  $$\int\sqrt{\h g}\Rh \f1\Boxh \Rh = \int\sqrt{\gh(\ze_1,\tau_1)}\int\sqrt{\gh(\ze_2,\tau_2)} R_1 G\qty(\{\ze_1,\tau_1\};\{\ze_2,\tau_2\}) R_2 $$  term for the geometry, \eqref{eq:thermal-bulk-Euc}. We get,
	\begin{equation}\label{eq:rboxr}\begin{aligned}
		\int\sqrt{\h g}\Rh \f1\Boxh \Rh = \frac{\be }{\pi  b^2 \de }  -\frac{2}{b^2} \log(\be/\de) +\frac{2\log(4\pi)-3}{b^2} -\frac{\pi  \de }{3 b^2 \be }
	    \end{aligned}\end{equation}
Again, the linearly divergent piece that appears above is cancelled by the contribution coming from the thermal \ads partition function.\footnote{It can be seen easily by doing a similar computation using the Green's function, \eq{eq:green-thermal-final},  on the thermal \ads geometry as discussed in \autoref{sec:Greens-funct}.}

	Thus, the total action is
	\begin{equation}\label{eq:part-func}
	   \boxed{ \log(Z) = - \be F =  -\frac{2}{b^2} \log(\be/\de) +\frac{2\log(4\pi)-3}{ b^2} + \frac{\de g}{2b^2 \, \be} {\color{red}} + \mathcal{O}(\de g^2)  }
	\end{equation}

\section{Comparison with field theory \label{sec:matching}}

In this section, we will make some remarks comparing our gravity dual we
discussed above and the SYK model.

\subsection{Hydrodynamics and a double scaling}

The gravity dual leads to the following low energy effective action
\begin{align}
S_{hydro} & = \f{\delta g}{2\pi b^2} \int
d\taut\ \qty{\ft(\taut),\taut }
\end{align}
while the SYK model has the following expression for the same
quantity \cite{Kitaev-talks:2015, Maldacena:2016hyu}
\begin{align}
S_{hydro} & = N \f{\al(q)}{{\mathcal{J}}} \int
d\tau\ \qty{\ft(\tau),\tau}
\end{align}
As we argued above, $\delta g$ plays the role of the explicit
symmetry breaking parameter $1/J$ in the SYK model. Further, the
classical limit in the bulk model corresponds to $b\to 0$, which,
therefore corresponds to the limit $N\to \infty$. Therefore, we
identify these quantities up to constants, thus:
\begin{align}
\f1{b^2}= c_1 N, \;  \delta g= c_2 \f1{{\mathcal{J}}}
\label{parameter-match}
\end{align}
For the two hydrodynamic expressions above to match, we need to have
$c_1 c_2 = \al(q)$. A $q$-dependence in the coefficients $c_1, c_2$
may appear strange; however, it may indicate the existence of a double
scaling in the theory. Note that at large $q$, $\al(q) a_0/q^2$
($a_0$=constant). A possible choice of the coefficients is $c_1=
\al(q)$, $c_2=1$. In this case, we are essentially identifying
\begin{align}
\f1{b^2}= a_0 N/q^2, \;  \delta g= c_2 \f1{{\mathcal{J}}}
\label{parameter-match-2}
\end{align} 
Thus, if we take the limit $N \to \infty$, {\it and} $q^2/N$ fixed
({\it cf.} \cite{Cotler:2016fpe} appendix B), the corresponding scaled
quantity corresponds to the bulk Newton's constant:
\[
q^2/N= a_0 b^2
\]

\subsection{Thermodynamics}

At low temperatures, the bulk partition function is given by
\eqref{eq:part-func}, with a divergence of the form
$\log(\be/\de)$. With the logarithmically divergent term we might typically be left with finite parts, say $P_0$, after
cancellation of the divergence. The low temperature partition function will
then be given by, ignoring subleading order terms in $\delta g/\delta$,
\begin{equation}
\label{eq:part-func-again}
	    \log(Z) = - \be F = \f1{b^2} \left[ \left(
                - 2 P_0 + \frac{4\log(4\pi)-5}{2}
              \right) + \frac{\delta g}{2\beta} + \mathcal{O}(\de
              g^2)\right]
\end{equation}
The corresponding expression in the SYK model is \cite{Maldacena:2016hyu,Cotler:2016fpe,Garcia-Garcia:2016mno}
\begin{equation}\label{eq:part-func-SYK}
	    \log(Z) = - \be F = N
\left[ \be {\mathcal{J}}\f1{q^2} + \f12 \log 2 - \f{\pi^2}{4 q^2}
+  \f1{\be {\mathcal{J}}}\f{\pi^2}{2 q^2} + O(\f1{q^4}) 
\right]  
	\end{equation}
It is then possible that by suitably adjusting the finite part $P_0$ and the
constant $c_2$ introduced above, one can
match the zero-temperature entropy and the low temperature specific
heat. The SYK zero-temperature entropy here does not seem to be
universal; however, in the double scaling limit mentioned above, the
$N/q^2$ term {\it is} universal.

A more detailed understanding of the low energy thermodynamics is
clearly desirable and is under investigation presently. 

\section{Discussion and open questions}\label{sec:discussion}

In this work, we arrive at a proposal for a gravity dual of the low
energy sector of SYK-type models from symmetry considerations, more
precisely from the fact that the coadjoint orbit action of the \emph{Diff}
 group is the Polyakov action \eq{eq:Liou-cov-bdy}. We solve the
classical equations of motion and find that the solutions are
parametrized by a large diffeomorphism together with a specific
conformal factor (value of the Liouville mode) representing a
non-normalizable deformation. We compute the on-shell action which
evaluates the classical contribution to $\log Z$. The computation
leads to a Schwarzian action for the low energy hydrodynamic modes and
a specific heat which is linear at low temperatures. Thus, the low
energy behaviour of our proposed gravity dual reproduces that of
SYK-type models.

We will end with some remarks about possible UV properties of the bulk
dual. Recall that in usual AdS/CFT, such as in the example of ${\cal
  N}=4$ SYM theory on $S^3 \times R$, states with spin $>2$ acquire
very large anomalous dimensions $\ga \sim (g_{YM}^2 N)^{1/4}$ at
strong coupling \footnote{Primary operators with spin $\leq 2$ retain
  $O(1)$ anomalous dimensions. These correspond to spherical harmonics
  of gravitons with $E \sim O(1)/R_{AdS}$.}  The energy grows as $E
\sim \ga/R_{AdS}$ and the corresponding bulk state is identified as a
string state with mass $m_s = (g_{YM}^2 N)^{1/4}/R_{AdS}$. This
corresponds to the fact that the UV completion of the gravity theory
is string theory in AdS. In case of SYK-type models, the anomalous
dimensions of operators with spin higher than two, which form an
approximately Regge trajectory, remain $O(1)$ even at strong coupling.
From the point of the bulk dual, the usual mass-dimension formula
(which follows by using the relation between the \ads Laplacian and
Casimir of \slr) implies $E \sim \Delta/R_{AdS}$ (in our model,
$R_{AdS}\sim 1/\sqrt{\mu}$, see \eq{eq:eom-trace}). If we wish to
identify the `Reggeons' with possible string states, this would imply
that the `string length' is of the same order as the AdS radius. It is
not clear what such a dual string theory of light strings is. On the
other hand, the spectrum of these massive modes suggests that it may
be possible to incorporate these states in our bulk dual by adding to
the Polyakov action \eq{eq:Liou-cov-bdy} an infinite number of matter
fields $\eta_r$ minimally coupled to the metric (see
\cite{Maldacena:2016upp} for related ideas), with masses $m_r$ given
in terms of the conformal dimensions $\Delta_r$. In such a scenario,
the Polyakov action \eq{eq:Liou-cov-bdy} would still continue to
represent the physics of the `Nambu-Goldstone' modes. The
full action will have the structure
\begin{equation}
S=  S_{cov}[g] +  S_{matter}[g,\{\eta_r\}]
\end{equation}
where $S_{cov}[g]$ is the Polyakov action, given by \eq{eq:Liou-cov-bdy}. The
matter action 
\begin{align}
S_{matter}[g,\{\eta_r\}] &=  \f1{2}  \int_{\Ga} \sqrt{g}
\left[\sum_r
\left(g^{\al\be}\del_\al \eta_r \del_\be \eta_r +  m_r^2 \eta_r^2\right)
+ ...\right]
\nonumber\\
 &=  \f1{2}  \int_{\Ga} \sqrt{\hat g}
\left[\sum_r
\left({\hat g}^{\al\be}\del_\al \eta_r \del_\be \eta_r +  
m_r^2 e^{2\phi} \eta_r^2\right)+ ... \right]
\label{guess-string}
\end{align}
where in the second step, we have used \eq{non-norm-f}. Note that
since the metric $\hat g$ contains the Nambu-Goldstone modes $f$ (see
\eqref{eq:AAds-class}), the above action automatically incorporates a
coupling between these modes and the higher mass modes $\eta_r$; this
fact plays an important role in computing the chaotic growth of the
out-of-time correlator. Using the action (8.1) we can derive the exponentially growing behaviour of the out-of-time ordered 4- point functions, $\lan \mathcal O(\tau)\mathcal O(0)\mathcal O(\tau)\mathcal O(0) \ran$ (where $\tau>0$), which gives the Lyapunov exponent, $2\pi/\beta$, consistent with the bound on chaos derived in \cite{Maldacena:2015waa}. Note also the appearance of the Liouville
factor in the mass term (this is to be contrasted with proposed bulk
duals based on Jackiw-Teitelboim models, e.g. in
\cite{Maldacena:2016upp}). This implies subleading correction to the
mass term proportional to $1/J$ (see \eq{non-norm-f}). However, as
shown in \cite{Maldacena:2016hyu,Polchinski:2016xgd} one doesn't need
to break the conformal symmetry explicitly to study the physics of
these excited states. In fact, the $1/J$ corrections for these states
are truly subleading. The terms in the ellipsis above denote
interaction terms, which are suppressed in large $N$ counting.
Whether the procedure of incorporating bulk fields outlined above can
be consistently extended to an interacting level with local
interactions in the bulk, of course, remains an open question.

\subsection*{Acknowledgments}

We would like to thank L. Alvarez-Gaume, T. Batista, S. Bhattacharya,
A. Dabholkar, A. Dhar, R. Gopakumar, D. Gross, D. Harlow, S. Jain,
J. Maldacena, S. Minwalla, R. Poojary, S. Sachdev, A. Sen, A. Sinha,
R. Soni, J. Sonner, D. Stanford, A. Strominger, S. Trivedi,
H. Verlinde and E. Witten. G.M. would like to thank participants of
the meetings {\it String Theory: Past and Present}, ICTS-TIFR,
Bangalore, 11-13 January 2017 \cite{GM:Talk}, for stimulating
discussions where part of this work was presented.  The work of G.M
and P.N. was supported in part by Infosys Endowment for the study of
the Quantum Structure of Space Time. S.R.W. would like to thank Theory
Division of CERN Geneva where part of this work was done. S.R.W would like to acknowledge the support of the Infosys Foundation Homi Bhabha Chair at ICTS. We thank
S. Minwalla for insightful comments on the original draft of this
paper.\\ We would like to thank D. Stanford and E. Witten for
illuminating correspondences on the first version.


\appendix
\appendixpage

\section{Some identities}
For $g_{\al\be} = e^{2\phi}\gh_{\al\be}$
\begin{subequations}\label{eq:identities}
    \begin{align}
	R &= e^{-2\phi} \pqty{\Rh-2\Boxh \phi} \\
	\Box &= e^{-2\phi} \Boxh \\
	g := \det(g_{\al\be}) &= e^{4\phi} \h g \\
	n_\mu &= e^\phi \h n_\mu \\
	n^\mu &= e^{-\phi} n^\mu \\
	\ga := \det{\ga_{\al\be}} &= e^{2\phi} \h \ga \\
	\sqrt{\ga} \Kb &= \sqrt{\h \ga} \pqty{\h \Kb + \h n^\mu \del_\mu \phi}
    \end{align}
\end{subequations}

\section{Green's function of Laplacian in \texorpdfstring{\ads}{AdS2}\label{sec:Greens-funct}}
Green's functions in hyperbolic spaces are well studied. Therefore, in this appendix, following \cite{Borthwick:book}, we only provide a quick review of some results that are important for this paper. In the Poincare half plane, $\rhp$, the Laplacian is given by,
\begin{equation}\label{eq:lap-rhp}
	\h \Box = \ze^2 \pqty{\del_\ze^2 + \del_\tau^2}
\end{equation}
We are interested in solving the Green's function equation,
\begin{equation}\label{eq:GreenEq}
	\h \Box G(\vec{x},\vec{x}') = \ze^2 \de^{(2)}(\vec{x}-\vec{x}')
\end{equation}
It is convenient to work with the coordinates, $z=\ze+i \tau, \zb=\ze-i\tau$. Geodesic distances between two points, $z,z'$, on $\rhp$ are given by,
\begin{equation}\label{eq:geo-dist}
	d(z,z') = \f1{\sqrt{4 \pi \mu}} \arccos\qty(1 + \frac{|z-z'|^2}{2 \Re[z]\Re[z']})
\end{equation}
Hyperbolic symmetry implies that the Green's function depends only on the geodesic distance, $G(z,z') = f(d)$. Switching to geodesic polar coordinates centered around $z'$, 
\begin{equation}\label{eq:geo-polar}
	ds^2 = dr^2 + \sinh^2(2\sqrt{\pi\mu} \; r) d\th^2
\end{equation}
In these coordinates, \eqref{eq:GreenEq} becomes,
\begin{equation}\label{eq:GreenEqGeoPol}
	\bqty{ \f1{\sinh(2\sqrt{\pi\mu} \; r)} \, \del_r \pqty{ \sinh(2\sqrt{\pi\mu} \; r) \del_r } } f(r) = \f{ \de(r) } { \sinh(2\sqrt{\pi\mu} \; r) }
\end{equation}
We regulate the above equation by first solving the resolvent for the operator $-\Box+4\pi\mu \; s(s-1)$, and then taking the limit, $s\to1$. Moreover, we first solve the homogeneous condition and then impose an appropriate condition on the discontinuity of the resolvent at origin to solve for the Green's function. The solution to the regulated homogeneous equation,
\begin{equation}\label{eq:homo}
	\bqty{ \f1{\sinh(2\sqrt{\pi\mu} \; r)} \, \del_r \pqty{ \sinh(2\sqrt{\pi\mu} \; r) \del_r } + s(s-1)} f_s(r) = 0
\end{equation}
is given by,
\begin{equation}\label{eq:sol-homo}
	f_s(r) = a_1 Q_{s-1}\pqty{ \cosh(2\sqrt{\pi\mu} \; r) } + a_2 P_{s-1} \pqty{ \cosh(2\sqrt{\pi\mu} \; r) } 
\end{equation}
where, $P_s, Q_s$ are Legendre functions of first and second kind respectively, and  $a_1,a_2$ are some constant of integrations. To fix the normalization and the discontinuity at the origin, we substitute \eqref{eq:sol-homo} into \eqref{eq:GreenEqGeoPol} and integrate on both sides. This fixes the solution for the resolvent to be,
\begin{equation}\label{eq:resolv}\begin{aligned}
	f_s(r) &= - \f1{2\pi} Q_{s-1} \pqty{ \cosh(2\sqrt{\pi \mu} r) } \\
	G_s(z,z') &= - \f1{2\pi} Q_{s-1} \pqty{ 1 + \frac{|z-z'|^2}{ 2 \Re[z] \Re[z']} } \\
	&= - \frac{\Ga(s)^2}{4\pi\Ga(2s)} \pqty{ 1 + \frac{|z-z'|^2}{ 4 \Re[z] \Re[z']}  }^{-s} {}_2F_1\pqty{s,s;2s; \pqty{ 1 + \frac{|z-z'|^2}{ 4 \Re[z] \Re[z']} }^{-1} }
\end{aligned}\end{equation}
Taking $s\to1$, the Green's function is given by,
\begin{equation}\label{eq:Green-Final}
	G(z,z') = \f1{4\pi} \log\qty[1 - \pqty{ 1 + \frac{|z-z'|^2}{ 4 \Re[z] \Re[z']} }^{-1}]
\end{equation}
In terms of the $\ze-\tau$ coordinates, this is,
\begin{equation}\label{eq:Green-Final-zeta-tau}
	G\qty(\{\ze_1,\tau_1\},\{\ze_2,\tau_2\}) = \f1{4\pi} \log\qty[\frac{ \qty(\ze_1-\ze_2)^2+\qty(\tau_1-\tau_2)^2 }{ \qty(\ze_1+\ze_2)^2+\qty(\tau_1-\tau_2)^2 }]
\end{equation}
The Green's function is quite instructive in this form. It is same as the flat space Green's function in 2-dimensions with an additional contribution coming from the `mirror charge' at $\{-\ze_2,\tau_2\}$. This is not surprising because \ads is Weyl scaled flat metric and hence has the same Green's function up to imposition of boundary conditions.

	\subsection{Green's function for thermal \texorpdfstring{\ads}{AdS2}}
	Thermal \ads is defined by periodic identification of $\tau$ coordinate over a length $\beta$. Thus the metric remains same as pure \ads and so does the Laplacian given in \eqref{eq:lap-rhp}. However, now the Green's function should be invariant under the shift $\De \tau = \tau_1-\tau_2 \to \De \tau + n \be$, with $n\in \mathbb{Z}$. This can be achieved by taking using the method of images,
	\begin{align}\label{eq:green-thermal}
		G_{thermal}\qty(\{\ze_1,\tau_1\},\{\ze_2,\tau_2\}) = \f1{4\pi} \sum_{n=-\infty}^{\infty}\log\qty[\frac{ \qty(\ze_1-\ze_2)^2+\qty(\De\tau+n\be)^2 }{ \qty(\ze_1+\ze_2)^2+\qty(\De\tau+n\be)^2 }]
	\end{align}
	This sum can be computed explicitly,
	\begin{align}\label{eq:green-thermal-final}
		G_{thermal}\qty(\{\ze_1,\tau_1\},\{\ze_2,\tau_2\}) = \f1{4\pi} \log \left[\frac{\cosh \left(\frac{2 \pi  (\ze_1-\ze_2)}{\beta }\right)-\cos \left(\frac{2 \pi  \Delta \tau }{\beta }\right)}{\cosh \left(\frac{2 \pi  (\ze_1+\ze_2)}{\beta }\right) - \cos \left(\frac{2 \pi  \Delta \tau }{\beta }\right)}\right]
	\end{align}
	This Green's function was used in the computations of the partition function in \autoref{sec:Thermodynamics} which was then subtracted from the partition function in black hole geometries discussed in that section.

\section{Variation of the induced gravity (Polyakov) action}\label{sec:var-Poly}
In this appendix we will study the exact variation of the Polyakov action, \eqref{eq:Liou-cov-bdy}. We haven't found a discussion of these covariant equations of motion anywhere in literature, we think that it might have been worked out personally, they haven't been presented in published form. Since the action is non-local, so are the equations of motion.\footnote{This also makes this section pretty ugly in terms of the equations.} While we won't be solving the equations in full generality, we show,
 \begin{enumerate} 
	\item That the diagonal part of the equations of motion are the same as the one we obtain for the \emph{Liouville mode}, $\phi$, in conformal gauge. These is the equation of motion that one obtains for Liouville field theory with a background metric, $\gh$.
	\item \ads and A\ads geometries that we have discussed in the paper satisfy the equations of motion.
	\item The most general solutions (\cite{Seiberg:1990eb,Witten:1976ck}) of the Liouville mode, $\phi$, in \ads background,
		\begin{equation*}
			\phi = \f12 \log\qty[(z+\zb)^2 \frac{\del g(z) \delb \gb(\zb)}{(g(z)+\gb(\zb))^2}]
		\end{equation*}
	obtain further constraints from the equations of motion. That is not surprising because the above solutions were obtained from solving only the Liouville equation. This also bodes well with the degree of freedom counting in 2d theory of gravity. These constraints force the solutions of $g(z), \gb(\zb)$ to be $\slc$ transformations of complex plane, \eqref{eq:Virasoro-constraints-text}. However, the boundary conditions reduce it to $\slr$ transformations, which are the isometries of the geometries that we are interested in. The remaining solutions that don't satisfy the boundary conditions are what we call \emph{non-normalizable} solutions.
	\item This exercise also justifies the boundary terms that we have introduced in \eqref{eq:Liou-cov-bdy} that we have argued are required for a well defined variational principle.
 \end{enumerate} 
We use following notations to avoid clutter in the forthcoming equations: 
\begin{subequations}\begin{align}
	&\int_\Ga^x \equiv \int_\Ga d^2x \sqrt{g(x)} \\
	&\int_{\del\Ga}^s \equiv \int_{\del\Ga} ds \sqrt{\ga(s)} \text{ where $s$ is the boundary coordinate} \\
	&G(x,y) \text{ is the Green's function of the Laplacian satisfying, } \Box^{(x)} G(x,y) = \f{\de^2(x-y)}{\sqrt{g(x)}} \label{eq:greensEq}\\
	&  \cdev_\mu^{(x)} \text{ denotes the covariant derivative with respect to variable }x
\end{align}\end{subequations}

\paragraph{Bulk Term} 
We start with varying the bulk term in \eqref{eq:Liou-cov-bdy}. 
\begin{equation*}
\begin{aligned}
	\de S_{cov}^{bulk}[g] &=  \f1{16\pi b^2} \int_\Ga \de \pqty{ \sqrt{g} \bqty{R\f1\Box R-16\pi\mu} }\\
	&=  \f1{16\pi b^2} \int_\Ga d^2x \int_\Ga d^2y \;  \de \bqty{ \sqrt{g(x)} \sqrt{g(y)} R(x) G(x,y) R(y) } +  \f1{16\pi b^2} \int_\Ga d^2x \; \de\bqty{\sqrt{g(x)}} \; \pqty{-16\pi\mu}\\
	&\hspace{-5pt}=  \f1{16\pi b^2} \int_\Ga d^2x \int_\Ga d^2y \; \pqty{ 2 \, \de \bqty{ \sqrt{g(x)} R(x)} \sqrt{g(y)} R(y)  G(x,y) + \sqrt{g(x)} R(x) \sqrt{g(y)} R(y) \; \de \bqty{ G(x,y)} } \\
	& \qquad +  \f1{16\pi b^2} \int_\Ga d^2x \; \de\bqty{\sqrt{g(x)}} \; \pqty{-16\pi\mu}
\end{aligned}\end{equation*}
Here, in the last equation on RHS we have used the symmetry of Green's function in $x-y$ coordinates to multiply the first term by 2. In the above equation, the first and last term are very easy to compute while the second term is slightly more non-trivial. The variations of the Ricci scalar and metric determinant are:
\begin{align*}
	\de \bqty{ \sqrt{g(x)}} &= -\f12  \sqrt{g(x)} g_{\mu\nu}(x) \de g^{\mu\nu}(x)\\
	\de \bqty{ R(x)} &= R_{\mu\nu} \de g^{\mu\nu} + \cdev_\mu v^\mu\\
	&\quad \text{where, } v^\si = g_{\mu\nu} \cdev^\si (\de g^{\mu\nu}) - \cdev_\al (\de g^{\al\si}) 
\end{align*}
Henceforth, we are dropping the overall factor of $1/16\pi b^2$ and will reinstate it at the end.
\begin{equation*}
\begin{aligned}
	\de S_{cov}^{bulk}[g] &=  2\int_\Ga^x \int_\Ga^y \pqty{ R_{\mu\nu}(x) -\f12  g_{\mu\nu}(x) R(x) } \de g^{\mu\nu}(x) \, R(y)  G(x,y) + 2\int_\Ga^x \int_\Ga^y \; \cdev^{(x)}_\mu v^\mu R(y)  G(x,y) \\
	& \qquad + \int_\Ga^x \int_\Ga^y R(x) R(y) \; \de \bqty{ G(x,y)}  +  \int_\Ga^x  8\pi\mu \; g_{\mu\nu}(x) \de g^{\mu\nu}(x)\\[10pt]
	&= 2\int_\Ga^x \int_\Ga^y \; \cdev^{(x)}_\si \bqty{g_{\mu\nu}(x) \cdev_{(x)} ^\si (\de g^{\mu\nu}(x)) - \cdev^{(x)}_\al (\de g^{\al\si}(x))} R(y)  G(x,y)  \\
	& \qquad + \int_\Ga^x \int_\Ga^y R(x) R(y) \; \de \bqty{ G(x,y)} +  \int_\Ga^x  8\pi\mu \; g_{\mu\nu}(x) \de g^{\mu\nu}(x)
\end{aligned}\end{equation*}
where, in the second line we have dropped the term containing Einstein tensor which is identically zero in 2 dimensions. Subsequently, we integrate by parts, keeping track of all the boundary terms that we pick in the process.
\begin{equation}\label{eq:bulk-var}\begin{aligned}
	\de S_{cov}^{bulk}[g] &= -2\int_\Ga^x \int_\Ga^y \; \bqty{ \cdev_{(x)} ^\si \qty(g_{\mu\nu}(x) \de g^{\mu\nu}(x)) - \cdev^{(x)}_\al (\de g^{\al\si}(x))} R(y)  \cdev^{(x)}_\si G(x,y) \\
	& \qquad  + 2\int_\Ga^y \int_\Ga^x \;  \cdev^{(x)}_\si \bqty{ v^\si(x) \, R(y)  G(x,y) } + \int_\Ga^x \int_\Ga^y R(x) R(y) \; \de \bqty{ G(x,y)} +  \int_\Ga^x  8\pi\mu \; g_{\mu\nu}(x) \de g^{\mu\nu}(x) \\[10pt]
	&= -2\int_\Ga^x \int_\Ga^y \; \bqty{ \cdev_{(x)} ^\si \qty(g_{\mu\nu}(x) \de g^{\mu\nu}(x)) - \cdev^{(x)}_\al (\de g^{\al\si}(x))} R(y)  \cdev^{(x)}_\si G(x,y) \\
	& \qquad  + 2\int_\Ga^y \int_{\del\Ga}^s \;  \nh_\si(s) v^\si(s) \; R(y)  G(x,y) + \int_\Ga^x \int_\Ga^y R(x) R(y) \; \de \bqty{ G(x,y)} +  \int_\Ga^x  8\pi\mu \; g_{\mu\nu}(x) \de g^{\mu\nu}(x)\\[10pt]
	&= -4\int_\Ga^y \int_{\del\Ga}^s \; \de\Kb \; R(y)  G(x,y) -2\int_\Ga^x \int_\Ga^y \; \cdev_{(x)} ^\si \qty(g_{\mu\nu}(x) \de g^{\mu\nu}(x) \; R(y)  \cdev^{(x)}_\si G(x,y) ) \\
	& \qquad + \int_\Ga^x \; g_{\mu\nu}(x) \de g^{\mu\nu}(x) \Big( 2R(x) +  8\pi\mu \Big) + 2\int_\Ga^x \int_\Ga^y \; \cdev^{(x)}_\al \pqty{ \de g^{\al\si}(x) \; R(y)  \cdev^{(x)}_\si G(x,y) }\\
	& \qquad  - 2\int_\Ga^x \int_\Ga^y \; \de g^{\al\si}(x) \cdev^{(x)}_\al\cdev^{(x)}_\si G(x,y) \; R(y)  + \int_\Ga^x \int_\Ga^y R(x) R(y) \; \de \bqty{ G(x,y)}
\end{aligned}\end{equation}
in the second line on RHS, we have used the Gauss's law to make the bulk integral into a surface integral. The first term in the second line is also the term that needs to be cancelled because it involves derivatives of variation of metric. Using $\nh^\si v_\si = -2 \de \Kb$, one can clearly see that this term is cancelled by the variation of second term in \eqref{eq:Liou-cov-bdy}. In the third line, we have used integration by parts in the second term of the second line. We have obtained two boundary terms in the process ($2^{nd}$ and $4^{th}$ term in the third line), both of which involve variation of the metric on the boundary, and under Dirichlet boundary condition, are zero. They will be dropped from here onwards.
\begin{equation}\label{eq:bulk-var-inter1}
\begin{aligned}
	\de S_{cov}^{bulk}[g] &= -4\int_\Ga^y \int_{\del\Ga}^s \; \de\Kb \; R(y)  G(x,y) + \int_\Ga^x \; g_{\mu\nu}(x) \de g^{\mu\nu}(x) \Big( 2R(x) +  8\pi\mu \Big) \\
	& \qquad  - 2\int_\Ga^x \int_\Ga^y \; \de g^{\al\si}(x) \cdev^{(x)}_\al\cdev^{(x)}_\si G(x,y) \; R(y)  + \int_\Ga^x \int_\Ga^y R(x) R(y) \; \de \bqty{ G(x,y)}
\end{aligned}\end{equation}
Now we embark upon the computation of $\de \bqty{G(x,y)}$. The Green's function in a curved background is defined in a covariant manner by \eqref{eq:greensEq}. Varying this equation with respect to metric,
\begin{equation}\begin{aligned}
	\de\Box^{(x)} \, G(x,y) + \Box^{(x)} \de G(x,y) &= \frac{1}{2 \sqrt{g(x)}} g_{\mu\nu}(x) \de g^{\mu\nu} (x) \; \de^2(x-y)\\
	\de G(x,y) &=  \frac{1}{2} g_{\mu\nu}(y) \de g^{\mu\nu} (y) \; G(x,y) - \int_\Ga^w G(x,w) \de\Box^{(w)} \, G(w,y) \label{eq:var-green}
\end{aligned}\end{equation}
Here in the second line we have integrated both sides with a Green's function. The action of Laplacian on a scalar is also given by, $\Box^{(x)} f(x) = \f1{\sqrt{g(x)}} \del_\mu \pqty{ \sqrt{g(x)} g^{\mu\nu} (x) \del_\nu f(x)}$. Thus the variation of the $\Box $ operator is,
\begin{align}
	\de \Box^{(x)} \; f(x) &= \f1{2}  g_{\mu\nu}(x) \de g^{\mu\nu} (x) \; \Box^{(x)} f(x) - \f1{2 \sqrt{g(x)}} \del_\mu \pqty{ \sqrt{g(x)} g_{\rho\si} (x) \de g^{\rho\si} (x)  g^{\mu\nu} (x) \del_\nu f(x)} \nonumber\\
	& \qquad+ \f1{\sqrt{g(x)}} \del_\mu \pqty{ \sqrt{g(x)} \de g^{\mu\nu} (x) \del_\nu f(x)} \nonumber\\
	&= - \f1{2} g^{\mu\nu} (x) \del_\nu f(x) \; \cdev_\mu \Big( g_{\rho\si} (x) \de g^{\rho\si} (x)  \Big) + \f1{\sqrt{g(x)}} \del_\mu \pqty{ \sqrt{g(x)} \de g^{\mu\nu} (x) \del_\nu f(x)} \nonumber\\
	&= - \f1{2} g^{\mu\nu} (x) \del_\nu f(x) \; \cdev_\mu \Big( g_{\rho\si} (x) \de g^{\rho\si} (x)  \Big) + \del_\mu \pqty{ \de g^{\mu\nu} (x)}  \; \del_\nu f(x) + \de g^{\mu\nu} (x) \;  \del_\mu{ \del_\nu f(x)} \nonumber\\
	& \qquad + \f1{2 g(x)} \del_\mu \pqty{g(x)} \; \de g^{\mu\nu} (x) \del_\nu f(x)\nonumber\\
	&= - \f1{2} g^{\mu\nu} (x) \del_\nu f(x) \; \cdev_\mu \Big( g_{\rho\si} (x) \de g^{\rho\si} (x)  \Big) + \del_\mu \pqty{ \de g^{\mu\nu} (x)}  \; \del_\nu f(x) + \de g^{\mu\nu} (x) \;  \del_\mu{ \del_\nu f(x)} \nonumber\\
	& \qquad + {\Ga^\si}_{\mu\si} \; \de g^{\mu\nu} (x) \del_\nu f(x)\nonumber\\
	\Rightarrow \de \Box^{(x)} \; f(x) &= - \f1{2} g^{\mu\nu} (x) \del_\nu f(x) \; \cdev_\mu \Big( g_{\rho\si} (x) \de g^{\rho\si} (x)  \Big) + \cdev_\mu \pqty{ \de g^{\mu\nu} (x)}  \; \del_\nu f(x) + \de g^{\mu\nu} (x) \;  \cdev_\mu{ \del_\nu f(x)} \label{eq:var-box}
\end{align}
We have used chain rule of differentiation to come from the first line on RHS to the second line. We have also changed the normal derivative acting on $g_{\rho\si} (x) \de g^{\rho\si} (x)$ into a covariant derivative because it is a scalar. In the fourth line we have used the identity, $\del_\mu g(x) = 2 g(x) {\Ga^\nu}_{\mu\nu}$. We have also converted some of the differentiations into covariant derivatives in last line. For our computations, the role of $f$ in the above computations is played by, $\int^y_\Ga \sqrt{g(y)} R(y) G(w,y)$. Using \eqref{eq:var-box} in \eqref{eq:var-green} and substituting back into last term of \eqref{eq:bulk-var-inter1},
\begin{equation*}
\begin{aligned}
	\int_\Ga^x \int_\Ga^y R(x) R(y) \; \de \bqty{ G(x,y)} &= \int_\Ga^x \int_\Ga^y R(x) R(y) \bqty{  \frac{1}{2} g_{\mu\nu}(y) \de g^{\mu\nu} (y) \; G(x,y) - \int_\Ga^w G(x,w) \de\Box^{(w)} \, G(w,y) }\\[10pt]
	&= \frac{1}{2} \int_\Ga^x \int_\Ga^y R(x) R(y) g_{\mu\nu}(y) \de g^{\mu\nu} (y) \; G(x,y) \\
	& + \f12 \int_\Ga^x \int_\Ga^y \int_\Ga^w R(x) R(y) G(x,w) \Bigg[ g^{\mu\nu} (w) \pdv{w^\nu} G(w,y) \; \cdev^w_\mu \Big( g_{\rho\si} (w) \de g^{\rho\si} (w)  \Big) \\
	& \qquad \ -2 \cdev^w_\mu \pqty{ \de g^{\mu\nu} (w)}  \; \pdv{w^\nu} G(w,y) -2 \de g^{\mu\nu} (w) \;  \cdev^w_\mu{ \pdv{w^\nu} G(w,y)}  \Bigg] \\[10pt]
	&= \frac{1}{2} \int_\Ga^x \int_\Ga^y R(x) R(y) g_{\mu\nu}(y) \de g^{\mu\nu} (y) \; G(x,y) \\
	& - \f12 \int_\Ga^x \int_\Ga^y \int_\Ga^w R(x) R(y) \bqty{  G(x,w) \; \Box^{(w)} G(w,y) \;  g_{\rho\si} (w) \de g^{\rho\si} (w)  } \\
	& - \f12 \int_\Ga^x \int_\Ga^y \int_\Ga^w R(x) R(y) \bqty{ g_{\rho\si} (w) \;  g^{\mu\nu} (w) \pdv{w^\mu} G(x,w) \pdv{w^\nu} G(w,y) }  \de g^{\rho\si} (w) \\
	& + \f12 \int_\Ga^x \int_\Ga^y \int_\Ga^w R(x) R(y) \cdev^w_\mu \bqty{ G(x,w) g^{\mu\nu} (w) \pdv{w^\nu} G(w,y) \;  g_{\rho\si} (w) \de g^{\rho\si} (w)  } \\
	& \qquad \ - \int_\Ga^x \int_\Ga^y \int_\Ga^w R(x) R(y) \cdev^w_\mu \bqty{ G(x,w) \de g^{\mu\nu} (w)  \; \pdv{w^\nu} G(w,y) } \\
	& \qquad \ + \int_\Ga^x \int_\Ga^y \int_\Ga^w R(x) R(y) \bqty{ \cdev^w_\mu G(x,w) \de g^{\mu\nu} (w)  \; \pdv{w^\nu} G(w,y) } \\
	& \qquad \ + \int_\Ga^x \int_\Ga^y \int_\Ga^w R(x) R(y) G(x,w) \, \de g^{\mu\nu} (w)  \; \cdev^w_\mu \pdv{w^\nu} G(w,y) \\
	& \qquad - \int_\Ga^x \int_\Ga^y \int_\Ga^w R(x) R(y) G(x,w)  \de g^{\mu\nu} (w) \;  \cdev^w_\mu{ \pdv{w^\nu} G(w,y)} 
\end{aligned}\end{equation*}
in the third line on RHS, the first two terms cancel between themselves, while the last two terms also cancel between themselves. The fourth and the fifth terms are total derivative terms that are essentially some boundary terms. These terms vanish since we are working with Dirichlet boundary conditions such that the Green's function vanishes on the boundary.
\begin{equation}
\begin{aligned}
	\int_\Ga^x \int_\Ga^y R(x) R(y) \; \de \bqty{ G(x,y)} &=\\
	& \hspace{-3cm} \int_\Ga^x \int_\Ga^y \int_\Ga^w R(x) R(y) \bqty{ \partiald{G(w,x)}{w^\mu} \partiald{G(w,y)}{w^\mu} - \f12 g_{\mu\nu}(w) g^{\al\be}(w) \partiald{G(w,x)}{w^\al} \partiald{G(w,y)}{w^\be}} \de g^{\mu\nu} (w)
\end{aligned}\end{equation}
Thus, the final expression of the variation of the bulk action is (with the reinstating of the overall $\f1{16\pi b^2}$ factor),
\begin{equation}\label{eq:bulk-var-final}
\begin{aligned}
	\de S_{cov}^{bulk}[g] &= - \f1 {4\pi b^2} \int_\Ga^y \int_{\del\Ga}^s \; \de\Kb \; R(y)  G(x,y) + \f1{16\pi b^2} \int_\Ga^x \; g_{\mu\nu}(w) \de g^{\mu\nu}(w) \Big( 2R(w) +  8\pi\mu \Big) \\
	& \qquad  -  \f1{8\pi b^2} \int_\Ga^x \int_\Ga^y \; \de g^{\al\si}(w) \cdev^{(w)}_\al\cdev^{(w)}_\si G(w,y) \; R(y) \\
	& \hspace{-5pt}+ \f1{16\pi b^2} \int_\Ga^x \int_\Ga^y \int_\Ga^w R(x) R(y) \bqty{ \partiald{G(w,x)}{w^\mu} \partiald{G(w,y)}{w^\mu} - \f12 g_{\mu\nu}(w) g^{\al\be}(w) \partiald{G(w,x)}{w^\al} \partiald{G(w,y)}{w^\be}} \de g^{\mu\nu} (w)
\end{aligned}\end{equation}

The bulk equations of motion are non local and given by:
\begin{equation}\label{eq:eom-bulk-exact}\begin{aligned}
	0 &= \f1{16\pi b^2} \Bigg( g_{\mu\nu}(w) \Big(2 R(w) + 8 \pi\mu \Big) +  \int_\Ga ^x \bqty{ - 2 \cdev_\mu^{(w)} \cdev_\nu^{(w)} G(w,x) R(x) } \\
	& \hspace{15pt}+  \int_\Ga ^x \int_\Ga ^y  \bqty{ \partiald{G(w,x)}{w^\mu} \partiald{G(w,y)}{w^\mu} - \f12 g_{\mu\nu}(w) g^{\al\be}(w) \partiald{G(w,x)}{w^\al} \partiald{G(w,y)}{w^\be}} R(x) R(y) \Bigg)
\end{aligned} \end{equation}
Now let us look at the trace part of the equations of motion. The last term in the above equation doesn't contribute in that case.
\begin{equation}\begin{aligned}
	0 &=  \Bigg( 2 \bqty{2 R(w) + 8 \pi\mu} +  \int_\Ga ^x \bqty{ - 2 \Box^{(w)} G(w,x) R(x) } \Bigg) \\
	&= R(w) + 8 \pi\mu 
\end{aligned}\end{equation}
In conformal gauge, where $g_{\mu\nu}(x) = e^{2\phi(x)} \h g_{\mu\nu}(x)$, this is same as, \eqref{eq:eom},
\begin{equation}\label{eom-bulk-trace}
	\Rh(x) - 2 \h \Box \phi(x) = - 8 \pi\mu e^{2\phi(x)}
\end{equation}
which is also the equation of motion for the Liouville mode $\phi$ with background metric $\gh$. In \ads background ($d\hat s^2 = \pqty{1/\pi\mu (z+\zb)^2} dz\,d\zb$), the most general solution of this equation is, \cite{Seiberg:1990eb,Witten:1976ck},
\begin{equation}\label{eq:sol-liou-mode}
	\phi = \f12 \log\qty[(z+\zb)^2 \frac{\del g(z) \delb \gb(\zb)}{(g(z)+\gb(\zb))^2}]
\end{equation}
where, in Euclidean space, $g(z),\gb(\zb)$ are complex function which are complex conjugate of each other. Equivalently, in Lorentzian space, they can be chosen to be two independent real functions.

Solving \eqref{eq:eom-bulk-exact} in full generality is a daunting task that we don't undertake. We show that \ads satisfies these equations of motion, and also provide an argument that A\ads geometries satisfy them too. Traceless part of \eqref{eq:eom-bulk-exact} is,
\begin{equation}\label{eq:eom-bulk-traceless}\begin{aligned}
	0 &=   \int_\Ga ^x \bqty{ - 2 \pqty{\cdev_\mu^{(w)} \cdev_\nu^{(w)} G(w,x) - \f12 g_{\mu\nu}(w) \Box^{(w)} G(w,x) } R(x) } \\
	& \hspace{15pt}+  \int_\Ga ^x \int_\Ga ^y  \bqty{ \partiald{G(w,x)}{w^\mu} \partiald{G(w,y)}{w^\mu} - \f12 g_{\mu\nu}(w) g^{\al\be}(w) \partiald{G(w,x)}{w^\al} \partiald{G(w,y)}{w^\be}} R(x) R(y) 
\end{aligned} \end{equation}
One way to check that \ads satisfies the on-shell equations of motion is to directly use the \eqref{eq:Green-Final} in the above expression and do the exact computation. However, it is much easier if we think of \ads as Weyl scaling of flat space, $g^{AdS}_{\al\be} = e^{2\Om} \eta_{\al\be}$, where for $\rhp$, $\Om = -\log(\sqrt{\pi \mu} (z+\zb)) = -\log(\sqrt{4\pi \mu} \ze)$. We use the formula for Ricci scalar, $R(x) = -2 e^{-2\Om} \Box_{flat} \Om(x)$ to write \eqref{eq:eom-bulk-traceless} as,
\begin{equation}\label{eq:eom-bulk-traceless-flat}\begin{aligned}
	0 &=   4\int_\Ga d^2x e^{2\Om} \bqty{ \pqty{\cdev_\mu^{(w)} \cdev_\nu^{(w)} G(w,x) - \f12 g_{\mu\nu}(w) \Box^{(w)} G(w,x) } \pqty{e^{-2\Om} \Box^{(x)}_{flat} \Om(x)} } \\
	& \hspace{15pt} +  4\int_\Ga d^2x \int_\Ga d^2y  e^{2\Om(x)}e^{2\Om(y)} \bqty{ \partiald{G(w,x)}{w^\mu} \partiald{G(w,y)}{w^\mu} - \f12 g_{\mu\nu}(w) g^{\al\be}(w) \partiald{G(w,x)}{w^\al} \partiald{G(w,y)}{w^\be}} \\
	& \hspace{3cm}\times \pqty{e^{-2\Om(x)} \Box^{(x)}_{flat} \Om(x)} \pqty{e^{-2\Om(y)} \Box^{(y)}_{flat} \Om(y)}\\[5pt]
	&=   4\int_\Ga d^2x \bqty{ \pqty{\cdev_\mu^{(w)} \cdev_\nu^{(w)} \Box^{(x)}_{flat}G(w,x) - \f12 g_{\mu\nu}(w) \Box^{(w)} \Box^{(x)}_{flat}G(w,x) } \Om(x)} \\
	& \hspace{15pt} +  4\int_\Ga d^2x \int_\Ga d^2y \Bigg[ \partiald{\pqty{\Box^{(x)}_{flat}  G(w,x)}}{w^\mu} \partiald{\pqty{\Box^{(y)}_{flat} G(w,y)}}{w^\mu} \\
	&\hspace{4cm} - \f12 g_{\mu\nu}(w) g^{\al\be}(w) \partiald{\pqty{\Box^{(x)}_{flat} G(w,x)}}{w^\al} \partiald{\pqty{\Box^{(y)}_{flat} G(w,y)}}{w^\be} \Bigg] \times \Om(x) {\Om(y)} \\[5pt]
	&=   4\bqty{ \cdev_\mu^{(x)} \cdev_\nu^{(x)} \Om(x) - \f12 g_{\mu\nu}(w) \Box^{(w)} \Om(x) } +  4 \Bigg[ \partiald{ \Om(x)}{w^\mu} \partiald{ \Om(y)}{w^\mu} - \f12 g_{\mu\nu}(x) g^{\al\be}(x) \partiald{ \Om(x)}{x^\al} \partiald{{\Om(y)}}{x^\be} \Bigg] \\
	& = 0
\end{aligned} \end{equation}
In the second line we have used integration by parts to shift $\Box_{flat}$ on the corresponding Green's functions; we have dropped the vanishing boundary terms on our way. We also use the fact discussed at the end of \autoref{sec:Greens-funct}, that the Green's function remain unchanged for the Weyl scaled metrics, upto impositions of boundary condition. In this case the boundary condition, $G\qty(\{\ze_,\tau\},\{0,\tau_2\})=0$, is imposed by adding a contribution of a `mirror charge' at a point reflected across the boundary. Thus the flat space Laplacian acting on this Green's function gives two $\delta$-functions, one each for the `original charge' and `mirror charge'.\footnote{The $\delta$-function is a flat space $\de$-function.} The $\delta$-function of the mirror charge lies outside the region of integration and hence doesn't contribute.

The equations of motion \eqref{eom-bulk-trace},\eqref{eq:eom-bulk-traceless} are covariant equations under diffeomorphisms. Thus they will also be satisfied for the class of geometries that we constructed in \autoref{sec:asymp-ads}.\\
We can still do slightly better and solve the equations of motion for Weyl scaled metrics around a given background. Around \ads background, from \eqref{eq:eom-bulk-traceless} we get following \emph{Virasoro constraints} for $\phi$,
\begin{equation}\label{eq:Virasoro-constraints-ads}
	4 \left(
		\begin{array}{cc}
 			\del^2\phi (z,\zb)- \pqty{\del\phi(z,\zb)}^2 + 2 \dfrac{\del\phi(z,\zb)}{z+\zb}& 0 \\[-10pt]
			 0 & \delb^2 \phi (z,\zb)- \pqty{\delb\phi(z,\zb)}^2 + 2 \dfrac{\delb\phi(z,\zb)}{z+\zb}
		\end{array}
	\right) = 0
\end{equation}
Solving \eqref{eom-bulk-trace} and \eqref{eq:Virasoro-constraints-ads} simultaneously, we get solutions of the type \eqref{eq:sol-liou-mode}, but with $g,\gb$ additionally restricted by the conditions,
\begin{equation}\label{eq:Virasoro-constraints}
	0=\left(
\begin{array}{cc}
 2\pqty{\dfrac{ g^{(3)}(z)}{g'(z)}-\dfrac32 \dfrac{g''(z)^2}{g'(z)^2} }& 0 \\
 0 &2\pqty{\dfrac{ \gb^{(3)}(z)}{\gb'(z)}-\dfrac32 \dfrac{\gb''(z)^2}{\gb'(z)^2} } \\
\end{array}
\right)
\end{equation}
which is basically the Schwarzian derivatives of $g(z)$ and $\gb(\zb)$. This restricts $g(z)$ to be of the form,
\begin{equation}
	g(z) = \frac{a z+ i b }  { i c z + d }
\end{equation}
for $a,b,c,d \in \mathbb{C}$, and $\gb(\zb)$ is its complex conjugate. Imposing the boundary condition, $g(z)+\gb(\zb)|_{z+\zb=0} = 0$ further restricts $a,b,c,d \in \real$. These precisely corresponds to the isometries of the geometries that we are considering. However, more general choice of these parameters gives us solutions that we call \emph{non-normalizable} in this paper. These solutions diverge as $1/\ze$ for small deviations around identity,
\begin{equation}\label{eq:small-def}\begin{aligned}
	a &= 1+ \de a \\
	b &= \de b \\
	c &= \de c \\
	d &= 1 - \de a \\
    \end{aligned}\end{equation}

\paragraph{Boundary Term} A similar analysis for the variation of boundary terms in \eqref{eq:Liou-cov-bdy} gives,
\begin{equation}\label{eq:bdy1-var}\begin{aligned}
	\de S_{cov}^{bdy}[g] &= \f1{16\pi b^2} \int_\Ga \de \pqty{ 4 \sqrt{\ga} \Kb\f1\Box R }\\
	&=\f1{4\pi b^2} \int_\Ga^x \int_{\del\Ga}^s \de \Kb(s) \ G(x,s) R(x) - \f1{4\pi b^2}  \int_\Ga^x \int_{\del\Ga} ^s \de g^{\mu\nu}(x) \bqty{ \cdev_\mu^{(x)} \cdev_\nu^{(x)} G(x,s) \Kb(s) } \\
	& \hspace{-5pt}+\f1{4\pi b^2}  \int_{\del\Ga} ^s \int_\Ga ^x \int_\Ga ^w \de g^{\mu\nu}(w) \bqty{ \partiald{G(w,x)}{w^\mu} \partiald{G(w,s)}{w^\mu} - \f12 g_{\mu\nu}(w) g^{\al\be}(w) \partiald{G(w,x)}{w^\al} \partiald{G(w,s)}{w^\be}} \Kb(s) R(x) \\
	& \quad- \f1{2\pi b^2} \int_{\del\Ga}^s \int_{\del\Ga}^{s'} \de \Kb(s) G(s,s') \Kb(s')
\end{aligned}\end{equation}
Note that the first term in RHS of \eqref{eq:bdy1-var} exactly cancels the last term in RHS of \eqref{eq:bulk-var}. Moreover, the last term in \eqref{eq:bdy1-var} exactly cancels the variation arising from the last term in \eqref{eq:Liou-cov-bdy}. Also in writing the above expressions we have made use of the fact that we are imposing Dirichlet boundary conditions on the metric, $\de g_{\mu\nu}|_{\del\Ga}=0$

\section{Analysing off-shell constraints}\label{app:off-shell}
In this section we demonstrate that the constraints coming from the traceless part of the equations of motion in the conformal gauge, {\it viz.} the
`Virasoro constraints' \eqref{virasoro-const}, do not permit any off-shell degrees of freedom apart from those representing large diffeomorphisms of \ads geometry. 

It is enough to carry out this analysis in absence of the large
diffeomorphisms, with $\widehat{ds^2}$ as in \eq{ads2}. The generalization
to \eq{eq:AAds-class} is obtained by applying the large diffeomorphism
\eq{eq:exact-large-diff}, in the manner explained in Section \ref{sec:asymp-ads}.

Simplifying the holomorphic part of the constraints,
\begin{align}
&\del^2\phi (z,\zb)- \pqty{\del\phi(z,\zb)}^2 + 2 \dfrac{\del\phi(z,\zb)}{z+\zb}
=0\nn \\
\Rightarrow &~ \del\pqty{(z+\zb)^2\ \del \pqty{ e^{-\phi(z,\zb) } }} = 0
\label{linear-vir} \\
\Rightarrow &~ \del \pqty{ e^{-\phi(z,\zb) } }  = \frac{ A(\zb) }{ (z+\zb)^2 }\nn \\
\Rightarrow &~ e^{-\phi(z,\zb) } = - \frac{ A(\zb) }{ (z+\zb) } + B(\zb)
\label{a-b}
\end{align}
Similarly, solving the anti-holomorphic part gives,
\begin{equation}
\begin{aligned}
 e^{-\phi(z,\zb) } = - \frac{ C(z) }{ (z+\zb) } + D(z) 
\label{c-d}
\end{aligned}
\end{equation}
In the above equations, the functions $A, B, C, D$ are arbitrary and
independent, to begin with, as they appear as ``constants' of
integration.  However, they must satisfy the requirement that the two
expressions \eq{a-b} and \eq{c-d} for the same quantity
$e^{-\phi(z,\zb)}$ must be (i) equal to each other and (ii) 
real. Assuming a general power series form of each of the functions,
we find that these two requirements can only be met if $A, C$ are
quadratic and $B, D$ are linear, and, in particular, of the
form
\begin{equation}\begin{aligned}
A(\zb) = \bfa \zb^2 +2 i \bfb \zb + \bfc, 
\quad B(\zb) = \bfa \zb + \bfd + i \bfb \\
C(z) = \bfa z^2 -2 i \bfb z + \bfc, \quad D(z) = \bfa z+ \bfd
- i \bfb,
\end{aligned}\end{equation}
leading to the following solution for the Liouville field,
\begin{align}
	&e^{-\phi(z,\zb)} = \frac{\bfa z {\zb}+
(\bfd+i {\bfb})z+(\bfd-i\bfb){\zb}-
\bfc}{z+{\zb}}
\nn \\
\Rightarrow & ~ \phi(z,\zb) = \frac12 \log\qty[ \frac{ (z+\zb)^2 }{ (
\bfa z {\zb}+(\bfd+i {\bfb})z+(\bfd-i\bfb){\zb}-\bfc)^2 }]
\label{vir-sol}
\end{align}
Here the constants $\bfa, \bfb,\bfc,\bfd$ are real. Out of these four,
only three are physical. The reason is that the Virasoro constraints,
expressed as in \eq{linear-vir} (and the similar, antiholomorphic
equation) are homogeneous linear equations in the variable
$e^{-\phi(z,\zb)}$, which implies that $e^{-\phi(z,\zb)} \to {\rm
  constant}\times e^{-\phi(z,\zb)} $ is a symmetry of the
equations. Hence, the constants $\bfa, \bfb,\bfc,\bfd$ are only
determined up to a (real) scale factor.

It is important to check that the solution \eq{vir-sol} of the
Virasoro constraints satisfies the equation of motion \eq{eq:eom}.
This can be done in two ways:\\ (i) By direct substitution of
\eq{vir-sol} into \eq{eq:eom}, we find that \eq{eq:eom} is satisfied
up to a term proportional to $\bfa \bfc+ (\bfd+i \bfb) (\bfd- i \bfb)
- 1$. By using the scale symmetry mentioned above, we can clearly make
this vanish, e.g. by treating $\bfa, \bfb$ and $\bfd$ as independent
variables and fixing $\bfc= \left(1 - (\bfd+i \bfb)( \bfd- i
\bfb)\right)/\bfa$ (this is equivalent to choosing a gauge).  \\ (ii)
Alternatively, one can match \eq{vir-sol} with the solution
\eq{eq:sol-liou-mode-text}. We find that the parameters of the two
solutions are related as follows
\begin{align}
\bar a c -  a \bar c = - i \bfa, \;
\bar b d -  b \bar d=- i \bfc, \;
\bar b c + a \bar d= \bfd - i \bfb
\label{bold-face}
\end{align}
The \slc\ conditions $ad+ bc=1$ translate to the condition 
\begin{align}
\bfa \bfc+
(\bfd+i \bfb) (\bfd- i \bfb) = 1,
\label{hyperboloid}
\end{align}
As mentioned before, on this surface \eq{vir-sol} solves the equation
of motion \eq{eq:eom}. Furthermore, in the analysis of
\eq{eq:sol-liou-mode-text}, we found that the \slr subgroup,
parameterized by real values of $a,b,c,d$, correspond to trivial
isometries of \ads, and did not generate a new solution; there is a
natural interpretation of this fact according to \eq{bold-face}: real
$a,b,c,d$ translate to $\bfa=\bfb = \bfc=0$, $\bfd=1$, leading to the
trivial solution $\phi=0$. Thus the variables $\bfa,\bfb,\bfc,\bfd$,
given by \eq{bold-face} actually parameterize the nontrivial coset
\slc/\slr. In fact, the hyperboloid \eq{hyperboloid} parameterizes 
this coset.

As mentioned before, the above analysis can be generalized to the case
of the reference metric \eq{eq:AAds-class} by applying the large
diffeomorphism \eq{eq:exact-large-diff} to the solution \eq{vir-sol}.

\paragraph{Conclusion:}
The Virasoro constraints completely fix the Liouville field $\phi$ (up
to three real constants). As explained in the text (see Section
\ref{sec:asymp-ads}), the three constants need to be fixed as boundary
conditions for the path integral (since they correspond to
non-normalizable deformations). Thus, there are no off-shell variables
({\it i.e.} variables appearing in the path integration) that come from
the Liouville field $\phi$. The only off-shell variables are
represented by the large diffeomorphisms as in \eq{hydro-z}.

\section{Exact computation of asymptotic \texorpdfstring{\ads}{AdS2} geometries}\label{app:asymp-ads-comp}
We use the knowledge of exact asymptotically AdS$_3$ geometries to
construct A\ads geometries. In AdS$_3$ the space of solutions of
spacetimes with constant negative curvature is given by,
\cite{Banados:1998gg, Roberts:2012aq},
\begin{equation}\label{eq:ads3-metric}\begin{aligned}
	ds^2 = L^2_{(AdS_3)} \pqty{\frac{d\ze^2 + 2 dx d\xb}{\ze^2}\, + \, L(x) dx^2 + \bar{L}(\xb) d\xb^2 - \frac{    \ze^2}{2} L(x)\bar{L}(\xb) dx d\xb }
\end{aligned}\end{equation}
where, $L(x),\,\bar{L}(\xb)$ are holomorphic and anti-holomorphic functions, and related to the holographic stress tensor, \cite{Balasubramanian:1999re}. In the above references it is discussed how following large diffeomorphisms generate the above class of geometries from the Poincare AdS$_3$ geometry ($ds^2 = L^2_{(AdS_3)}(du^2+2dy d\yb)/u^2$),
\begin{equation}\label{eq:ads3-large-diff}\begin{aligned}
	y &= f(x) + \frac{2 \ze^2 f'(x)^2 \fb''(\xb)}{8f'(x)\fb'(\xb)-\ze^2 f''(x)\fb''(\xb)}\\
	y &= \fb(\xb) + \frac{2 \ze^2 \fb'(\xb)^2 f''(x)}{8f'(x)\fb'(\xb)-\ze^2 f''(x)\fb''(\xb)}\\
	u &= \ze \frac{\pqty{4 f'(x) \fb'(\xb)}^{3/2}}{8f'(x)\fb'(\xb)-\ze^2 f''(x)\fb''(\xb)}
\end{aligned}\end{equation}
In general, in any arbitrary dimensions, it is not difficult to solve for the asymptotic Killing vectors for any spacetime. The special feature of AdS$_3$ is the fact that these infinitesimal diffeomorphisms can be integrated to non-linear order. \ads being a more constrained geometry, also enjoys this same feature. Here, we use the known results of the exact non-linear diffeomorphisms in AdS$_3$ to construct the class of asymptotic \ads solutions. In Cartesian coordinates, $y=(x+i \tau)/\sqrt 2, \yb = (x-i \tau)/\sqrt 2$, one can obtain \ads as a reduction of AdS$_3$ b    y restricting to $x=0$ slice. Restricting ourselves to those transformations that keeps this \ads slice invariant,     i.e., for $f(x)+\fb(\xb)|_{x+\xb=0}=0$, we find that the following coordinate transformations are precisely those which generate large diffeomorphisms in \ads, \eqref{eq:exact-large-diff}, while keeping us within Fefferman-Graham gauge,
\begin{equation*}
     \taut = f(\tau) - \frac{2 \ze^2 f''(\tau) f'(\tau)^2}{4 f'(\tau)^2 + \ze^2 f''(\tau)^2}, \quad \zet = \frac{4 \  \ze f'(\tau)^3}{4 f'(\tau)^2 + \ze^2 f''(\tau)^2}
\end{equation*}
These transformations map the \ads metrics, $ds^2 = (d\zet^2 + d\taut^2)/(4\pi\mu \, \zet^2)$ to A\ads geometries,     $ds^2 = \qty(d\ze^2 + \qty(1-\frac{\ze^2}{2} \qty{f(\tau),\tau})^2 d\tau^2)/(4\pi\mu \, \ze^2).$

\section{\label{sec:quantum-corr}Quantum corrections to the classical action}
In this section we discuss the issue of gauge fixing the action \eqref{eq:Liou-cov-bdy}. The idea and Faddeev-Popov procedure to arrive at the same: We introduce a functional delta-function in our path integral using the Faddeev-Popov prescription. The corresponding determinant is then written in terms of fermionic ghosts, which gives rise to new ghost-graviton interaction vertices.\footnote{In the subsequent discussion in this particular appendix, we call the $f$ degree of freedom of \autoref{sec:asymp-ads} corresponding to large diffeomorphisms of \ads as `gravitons'.}

The Faddeev-Popov determinant is defined in terms of the gauge-fixing $\de$-function as follows,
\begin{equation}\label{eq:delta-def}
	1 = \De_{FP}\Big[ \gh[f(\tau)],\phi \Big] \times \int [\cD \ep^{(s)}] [\cD \phi] [\cD f(\tau)] \; \de\qty( g^{\e^{(s)}} - e^{2\phi} \gh[f(\tau)] ) \times \de\qty(\ep^{(s)}(z_1)) \, \de\qty(\ep^{(s)}(z_2)) \, \de\qty(\ep^{(s)}(z_3))
\end{equation}
Here, we are denoting the small diffeomorphisms (these are the gauge-symmetry of the theory) by $\ep^{(s)}$. In the subsequent discussion we will drop the $(s)$ superscript to conciseness. $\phi$ denotes the Weyl degree of freedom and will eventually become the Liouville mode. Since our theory is not Weyl-invariant, unlike in String theory, we don't factor out these degrees of freedom. Finally, $f(\tau)$ denotes the degree of freedom due to large diffeomorphisms that is discussed in \autoref{sec:asymp-ads}. We are gauge fixing (using only small diffeomorphisms) an arbitrary metric to a metric that is Weyl equivalent to metrics \eqref{eq:AAds-class}. This procedure will give us the Jacobian corresponding to change of integration `variable' from $[\cD g]$ to $[\cD \phi] [\cD f(\tau)]$. The $\de$-functions have been included in the above expression to fix the residual gauge symmetry corresponding to our gauge choice. This is precisely the $\slr$ isometry of the geometries \eqref{eq:AAds-class}, and hence we choose to fix three arbitrary points in the interior of A\ads geometries.\\The path integral that we are interested in computing is formally written as,
\[
	Z = \int \frac{[\cD g] }{ V_{ \ep } } e^{-S[g]}, \quad V_{ \ep } \text{ is the volume of the symmetry group}
\]
and inserting \eqref{eq:delta-def} into this path integral, we get,
\def\gt{\tilde{g}}
\begin{align}
	Z &= \int \frac{[\cD g] [\cD \ep] [\cD \phi] [\cD f(\tau)]}{ V_{ \ep } } \times \De_{FP}\Big[ \gh[f(\tau)],\phi \Big] \; \de\qty( g^{\e} - e^{2\phi} \gh[f(\tau)] ) \times e^{-S[g]} \times \pqty{\de\text{-functions}} \nonumber \\
	&= \int \frac{[\cD \gt] [\cD \ep] [\cD \phi] [\cD f(\tau)]}{ V_{ \ep } } \times \De_{FP}\Big[ \gh[f(\tau)],\phi \Big] \; \de\qty( \gt - e^{2\phi} \gh[f(\tau)] ) \times e^{-S[\gt]} \times \pqty{\de\text{-functions}} \nonumber \\
	&= \int \frac{[\cD \ep] [\cD \phi] [\cD f(\tau)]}{ V_{ \ep } } \times \De_{FP}\Big[ \gh[f(\tau)],\phi \Big] \times e^{-S[e^{2\phi} \gh[f(\tau)]]} \times \pqty{\de\text{-functions}} \nonumber \\
	&= \int [\cD \phi] [\cD f(\tau)] \times \De_{FP}\Big[ \gh[f(\tau)],\phi \Big] \times e^{-S[e^{2\phi} \gh[f(\tau)]]} \times \pqty{\de\text{-functions}}
\end{align}
In the second line on RHS, we have changed integration `variables' from $\cD g$ to $\cD\gt$, where, $g =: \tilde{g}^{\ep^{-1}}$ and used the fact that action and measure are both gauge invariant. In the third line we have integrated over the metric degrees of freedom using the $\de$-function. In the last line we have used the fact that the integrand of third line doesn't depend on $\ep$ anymore, integration over which simply gives us the volume of the symmetry group.\\
Faddeev Popov determinant can be easily written in terms of the $b$ and $c$ ghosts as,
\def\Ph{\hat{P}}
\begin{align}\label{eq:FP-det}
	\De_{FP}\Big[ \gh[f(\tau)],\phi \Big]  = \int \cD c_\al \; \cD b^{\al\be} \; \cD \Lfg_\al \exp[-\qty(b^{\al\be} (\Ph c)_{\al\be} - b^{\al\be} (\Ph \Lfg)_{\al\be})] \times \bqty{\frac{c(z_1) c(z_2) c(z_3)}{(z_1-z_2)(z_2-z_3)(z_3-z_1)}}
\end{align}
here, $c$-insertions are equivalent to the $\de$-functions appearing in the previous expressions. $b^{\al\be}$ is a symmetric-traceless tensor, and thus has only 2 degrees of freedom. We have defined operator $\Ph$ such that,
\def\fcdev{{}^{(f)}\cdev}
\[
	(\Ph x)_{\al\be} :=\fcdev_{(\al} x_{\be)} - (\fcdev \cdot x) \; \gh[f(\tau)]_{\al\be}
\]
$\fcdev$ is the covariant derivative w.r.t geometries in \eqref{eq:AAds-class}. $\Lfg$ is defined in terms of the fermionized large diffeomorphisms \eqref{eq:asymp-kill-vec} as,
\def\dLfg{\mathfrak{k}}
\[
	\Lfg = \begin{pmatrix} \ze \ \dLfg'(\tau) \\[-10pt] \dLfg(\tau) - \f{\ze^2}2 \dLfg''(\tau) \end{pmatrix}
\]
where again, $\dLfg$ are the fermionized ghost counter-part of the field appearing in \eqref{eq:asymp-kill-vec}. The above action can be expanded and written in terms of the components:
\begin{align}
	\Ph\Lfg = \left(
\begin{array}{cc}
 \substack{ \frac{1}{{\left(\{f(\tau),\tau\}\zeta ^2-2\right)^3}} \Big[ -2 \zeta ^4 \, \del_\tau \big( \{f(\tau),\tau\} \big)\dLfg''(\tau ) \\
 +4 \zeta ^2\, \del_\tau \big( \{f(\tau),\tau\} \big)\dLfg(\tau )+2 \zeta ^2\, \left(\{f(\tau),\tau\}\zeta ^2-2\right) \dLfg^{(3)}(\tau )\\
 +\left(\{f(\tau),\tau\}\zeta ^2-2\right) \left(\{f(\tau),\tau\}\left(\{f(\tau),\tau\}\zeta ^2-8\right) \zeta ^2+8\right) \dLfg'(\tau ) \Big] } &
 \frac{\left(\zeta ^2 \{f(\tau),\tau\}+2\right) \left(\zeta ^2 \dLfg''(\tau )-2 \dLfg(\tau )\right)}{\zeta  \left(\zeta ^2 \{f(\tau),\tau\}-2\right)} \\[30pt]
 \frac{\left(\zeta ^2 \{f(\tau),\tau\}+2\right) \left(\zeta ^2 \dLfg''(\tau )-2 \dLfg(\tau )\right)}{\zeta  \left(\zeta ^2 \{f(\tau),\tau\}-2\right)}
 & \substack{ \frac{1}{4 \{f(\tau),\tau\}\zeta ^2-8} \Big[ 16 \dLfg'(\tau ) + 2\ze^4 \del_\tau \big( \{f(\tau),\tau\} \big)\dLfg''(\tau ) \\
 - \ze^2 \{f(\tau),\tau\}\left(\{f(\tau),\tau\}\zeta ^2-6\right) \left(\{f(\tau),\tau\}\zeta ^2-4\right) \dLfg'(\tau )\\
 -4 \ze^2  \del_\tau \big( \{f(\tau),\tau\} \big)\dLfg(\tau )- 2 \ze^2 \left(\{f(\tau),\tau\}\zeta ^2-2\right) \dLfg^{(3)}(\tau ) \Big] }\\
\end{array}
\right)
\end{align}

\begin{align}
	\Ph c = \left(
\begin{array}{cc}
 \substack{ \frac{1}{\zeta  \left(\{f(\tau),\tau\} \zeta ^2-2\right)^3} \Big[ 4 \zeta ^3 \, \del_\tau \big( \{f(\tau),\tau\} \big) c_\tau(\zeta ,\tau ) \\
 -4 \zeta \, \left(\{f(\tau),\tau\} \zeta ^2-2\right) c_\tau^{(0,1)}(\zeta ,\tau ) \\
 + \zeta \, \left(\{f(\tau),\tau\} \zeta ^2-2\right)^3 c_\ze^{(1,0)}(\zeta ,\tau ) \\
 -4 \left(\{f(\tau),\tau\} \zeta ^2-2\right)^2 c_\ze(\zeta ,\tau ) \Big] } &
 \substack{ \left(c_\ze^{(0,1)}(\zeta ,\tau )+c_\tau^{(1,0)}(\zeta ,\tau )\right)\\
 -2\frac{\left(\zeta ^2 \{f(\tau),\tau\}+2\right) c_\tau(\zeta ,\tau )}{\zeta  \left(\zeta ^2 \{f(\tau),\tau\}-2\right)} } \\[30pt]
\substack{ \left(c_\ze^{(0,1)}(\zeta ,\tau )+c_\tau^{(1,0)}(\zeta ,\tau )\right)\\
 -2\frac{\left(\zeta ^2 \{f(\tau),\tau\}+2\right) c_\tau(\zeta ,\tau )}{\zeta  \left(\zeta ^2 \{f(\tau),\tau\}-2\right)} } 
 & \substack{ \frac{1}{4 \zeta  \left(\{f(\tau),\tau\} \zeta ^2-2\right)} \Big[ -4 \zeta ^3\, \del_\tau \big( \{f(\tau),\tau\} \big) c_\tau(\zeta ,\tau ) \\
 +4 \zeta\, \left(\{f(\tau),\tau\} \zeta ^2-2\right) c_\tau^{(0,1)}(\zeta ,\tau ) \\
 +4 \left(\{f(\tau),\tau\} \zeta ^2-2\right)^2 c_\ze(\zeta ,\tau )  \\
- \zeta\,\left(\{f(\tau),\tau\} \zeta ^2-2\right)^3 c_\ze^{(1,0)}(\zeta ,\tau ) \Big] }\\
\end{array}
\right)
\end{align}

\section{Weyl anomaly in manifolds with boundary}\label{app:Wess-Zumino}
In this section we compute the most general boundary term for Weyl anomaly in 2-dimensions on a manifold with a boundary allowed by the Wess-Zumino consistency condition. Let us start with the variation of \eqref{eq:Liou-cov-bdy} under a Weyl transformation,
\begin{align}
	\de_W S_{cov} &= -\frac1{4\pi b^2} \int\limits_\Ga \!\!\! \sqrt{g} \pqty{ R  + 8\pi\mu }\de \om + \frac1{4\pi b^2} \int\limits_{\del\Ga} \!\!\! \sqrt{\ga(s)} \int\limits_\Ga\!\!\! \sqrt{g(x)}\ \de\om(s) R(x)\ \nh^\mu(s) \pdv{y^\mu} G(x,y)|_{y=s} \nonumber \\
	&\quad + \frac1{2\pi b^2}  \int\limits_{\del\Ga} \!\!\! \sqrt{\ga(s_1)} \int\limits_{\del\Ga}\!\!\! \sqrt{\ga(s_2)}\ \de\om(s_2) \Kb(s_1) \ \nh^\mu(s_2) \pdv{y^\mu} G(s_1,y)|_{y=s_2} - \frac1{2\pi b^2} \int\limits_{\del\Ga}\!\!\! \sqrt{\ga(s)}\ \Kb(s) \de\om(s)
\end{align}
Under a second Weyl transformation
\begin{align}
	\de_{W_2} \pqty{ \de_{W_1} S_{cov} } &= -\frac1{2\pi b^2} \int_{\del\Ga} \sqrt{\ga(s)}\; \del^\mu \de\om(s)\; \del_\mu \de\om_2(s) - \frac{4\mu}{b^2} \int_\Ga\sqrt{g}\; \de\om_2(x) \; \de\om_1(x) \nonumber \\
	&\quad +\frac1{2\pi b^2} \int\limits_{\del\Ga}\sqrt{\ga(s_1)} \int\limits_{\del\Ga}\sqrt{\ga(s_2) } \; \de\om_1(s_1) \de\om_2(s_2) \; \nh^\mu(s_1) \nh^\nu(s_2) \; \del_\mu\del_\nu G(s_1,s_2)
\end{align}
therefore, since all the terms in the above equation are symmetric in $\de\om_1$ and $\de\om_2$, we have, 
\begin{align}
	\de_{W_2} \pqty{ \de_{W_1} S_{cov} } - \de_{W_1} \pqty{ \de_{W_2} S_{cov} } &= 0
\end{align}
Thus the boundary terms that we have introduced are consistent with the Wess-Zumino conditions. All the boundary terms that we have introduced are consistent with the general analysis in \cite{book:Polchinski}.

\enlargethispage{10pt}
\bibliographystyle{JHEP} 
\bibliography{1g} 

\end{document}